\documentclass[aps,prb,twocolumn,superscriptaddress,floatfix,letterpaper]{revtex4-1}

\usepackage[inline]{enumitem}
\usepackage{tikz}
\usetikzlibrary{arrows,decorations.markings}
\usepackage[export]{adjustbox}
\usepackage{float}
\usepackage[normalem]{ulem}
\usetikzlibrary{decorations.pathreplacing}

 \usepackage{pdftricks2}
 \usepackage{epsfig}
 \usepackage{pst-grad} % For gradients
 \usepackage{pst-plot} % For axes

\newcommand{\Rep}{{\cR\text{ep}}}
\newcommand{\ZZ}{\mathbb{Z}}
\newcommand{\fh}{\frac{1}{2}}
\newcommand{\nn}{\nonumber}
\renewcommand{\bar}{\overline}

\newcommand{\tAw}[3]{{#1}_{e_{#2#3} }^{v_{#2}v_{#3}}}

\newcommand{\tC}[5]{{#1}_{v_{#2}v_{#3}v_{#4}v_{#5};
\phi_{#2#3#5} \phi_{#3#4#5}
}^{
e_{#2#3} e_{#2#4} e_{#2#5} e_{#3#4} e_{#3#5} e_{#4#5};
\phi_{#2#3#4} \phi_{#2#4#5}
}}

\begin{document}

\begin{titlepage}
\title{ Non-invertible anomalies and\\
 mapping-class-group transformation of anomalous partition functions}

\author{Wenjie Ji}
\affiliation{Department of Physics, Massachusetts Institute of
Technology, Cambridge, Massachusetts 02139, USA}

\author{Xiao-Gang Wen}
\affiliation{Department of Physics, Massachusetts Institute of
Technology, Cambridge, Massachusetts 02139, USA}

\begin{abstract} 
Recently, it was realized that anomalies can be completely classified by
topological orders, symmetry protected topological orders, and symmetry
enriched topological orders in one higher dimension.  The anomalies that people
used to study are invertible anomalies that correspond to invertible
topological orders and/or symmetry protected topological orders in one higher
dimension.  In this paper, we introduce a notion of non-invertible anomaly,
which describes the boundary of generic topological order.  A key feature of
non-invertible anomaly is that it has several partition functions.  Under the
mapping class group transformation of space-time, those partition functions
transform in a certain way characterized by the data of the corresponding
topological order in one higher dimension.  In fact, the anomalous  partition
functions transform in the same way as the degenerate ground states of the
corresponding topological order in one higher dimension.  This general theory
of non-invertible anomaly may have wide applications.  As an example, we show
that the irreducible gapless boundary of 2+1D double-semion topological
order must have central charge $c=\bar c \geq \frac{25}{28}$.

\end{abstract}


\maketitle

\end{titlepage}

{\small \setcounter{tocdepth}{1} \tableofcontents }

\section{Introduction}

A classical field theory described by an action may have a gauge symmetry if
the action is gauge invariant. The corresponding theory is called a classical
gauge theory.  A gauge anomaly is an obstruction to quantize the classical
gauge theory, since the path integral measure may not be gauge
invariant.\cite{A6926,BJ6947} Similarly, a classical action may have a
diffeomorphism invariance.  Then a gravitational anomaly is an obstruction to
have a diffeomorphic invariant path integral.\cite{W8597} So the standard point
of view of anomaly corresponds to the obstruction to go from classical theory
to quantum theory.  This kind of gauge anomaly and gravitational anomaly are
always \emph{invertible}, \ie can be canceled by another anomalous theory.  The
examples include 1+1D $U(1)$-gauged chiral fermion theory
\begin{align}
 S = \int \dd x\dd t\ \psi^\dag (\prt_t+\ii A_t -\prt_x -\ii A_x)\psi,
\end{align}
which has both perturbative $U(1)$ gauge anomaly and perturbative gravitational
anomaly.  We like to remark that such defined anomaly is not a property of
physical systems, but a property of a formalism trying to convert a classical
theory to a quantum theory.

There is another invertible anomaly -- 't Hooft anomaly, that can be defined
within a quantum system with a global symmetry, and is a property of physical
systems.  It is not an  obstruction to go from classical theory to quantum
theory,  but rather an obstruction to gauge a global symmetry within a quantum
system.\cite{H8035}  It is quite amazing that the obstruction to quantize a
classical gauge theory (gauge anomaly) is closely related to the obstruction to
gauge a global symmetry within a quantum system. 

\begin{figure}[tb]
\begin{center}
\includegraphics[scale=0.6]{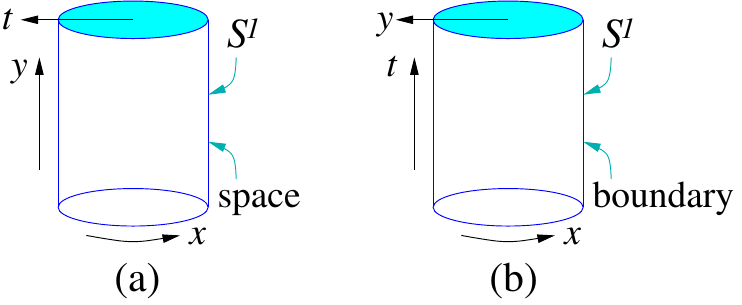} 
\end{center}
%Fig. 1
\caption{ 
(a) A particular time $t$ evolution produces a particular ground state in the
degenerate ground state subspace on the space $S^1\times S^1$.  (b) A
particular extension of a space-time $S^1\times S^1$ as the boundary of a bulk
$D^2\times S^1$ produces a particular anomalous partition function in the
vector space of partition functions on space-time $S^1\times S^1$ (\ie the
boundary).
}
\label{BdryGrnd}
\end{figure}

Motivated by some early results,\cite{CH8527,RML1204} in recent years, we
started to have a new understanding of anomaly as a physical property of
quantum systems:\cite{W1313,KW1458,KZ150201690} \frmbox{ The anomaly in a
theory directly corresponds to the topological order\cite{W8987,W9039} and/or
symmetry protected topological (SPT) order\cite{GW0931,CLW1141,CGL1314} (with
on-site symmetry) in one higher dimension. Such an anomalous theory is realized
by a boundary of the corresponding topological order and/or SPT order (see Fig.
\ref{BdryGrnd}b).}  So an anomaly is nothing but a topological order and/or a
SPT order in one higher dimension, and the anomalies can be classified via the
classification of topological orders and SPT orders in one higher
dimension.\cite{W1313,KT14030617}  The boundary of topological order realizes
all possible gravitational anomaly, and the boundary of SPT order realizes all
possible 't Hooft anomaly and mixed gravity/'t Hooft anomaly.  This point of
viewed of anomaly plus Atiyah formulation of topological quantum field
theory\cite{A8875} allow us to develop a general theory of
anomaly.\cite{W1313,KW1458,KZ150201690}

But the anomaly from this new point of view is not the same as the previously
defined anomaly before 2013, and  is more general.  This is because topological
orders are usually not invertible.\cite{KW1458,F1478} Hence, the anomaly
realized by the boundary of topological orders may be non-invertible as well
(\ie cannot be canceled by any other anomaly).  In contrast, the standard
anomalies are always invertible.  Thus the standard anomalies are classified by
invertible topological orders and/or SPT orders in one higher dimension, and
are realized by the boundary of the invertible topological orders and/or the
SPT orders.  But there are more general topological orders, which are not
invertible.  Those non-invertible topological orders will give rise to a new
kind of gravitational anomalies on the boundary, which will be called
non-invertible anomaly. For example, the chiral conformal field theory (CFT) on the boundary of a
generic Chern-Simons theory is an example of non-invertible
anomaly.\cite{W8951} 

We like to mention that, in addition to the above general point of view
proposed in \Ref{W1313,KW1458,KZ150201690} that include non-invertible
anomalies, \Ref{FV14095723,M14107442} also proposed a similar general point of
view based mathematical category theory and cobordism theory.  In particular,
those general points of view suggest that the partition function becomes a
vector in a vector space for a theory with non-invertible anomaly, as suggested
in \Ref{A8875}.  In fact, the vector space that contains partition functions
of an anomalous theory  can be identified with the degenerate ground state
subspace\cite{W8987,W9039} (see Fig.  \ref{BdryGrnd}a) of the topological order
in one higher dimension that characterizes the anomaly.  This is because if we
regard a space direction as time direction, the space manifold can be viewed as
a boundary of space-time, and the ground degenerate subspace becomes the vector
space of partition functions (see Fig.
\ref{BdryGrnd}).\cite{KW1458,WW180109938}

In this paper, we will study some simplest non-invertible anomalies -- bosonic
global gravitational anomalies in 1+1D which correspond to a 2+1D bosonic
topological order.  We will first give a general discussion, in particular the
physical meaning of ``partition function'' as a vector in a vector space.  Then
we will discuss some examples of 1+1D bosonic theories with non-invertible
gravitational anomalies that correspond to:
\begin{enumerate}
\item 2+1D bosonic $Z_2$ topological order\cite{RS9173,W9164} (\ie the
topological order described by the $Z_2$-gauge theory).  
\item 2+1D bosonic double-semion (DS) topological order.\cite{FNS0428,LW0510}
\item 2+1D bosonic  single-semion (SS) topological order (\ie $\nu=1/2$ quantum Hall state).\cite{L8395}
\item 2+1D bosonic Fibonacci topological order.\cite{FNS0428,LW0510}
\end{enumerate}

We will also discuss an application of invertible and non-invertible anomalies.
There is a general belief that a gapless CFT has a
partition function that is invariant under mapping class group (MCG)
transformations of the space-time (the modular transformations for
2-dimensional space-time), provided that the CFT can be put on a lattice.
Being able to put a CFT on a lattice is nothing but the anomaly-free condition.
This suggests that the MCG invariance of the partition function corresponds to
the anomaly-free condition.  So an anomalous CFT will have a partition function
which is not MCG invariant, but \emph{MCG covariant}.\footnote{The partition
functions of a fermionic CFT on a $d+1$D lattice are MCG covariant in a special
way, so that the sum of the partition functions are still invariant under the
fermionic MCG. For example, in 1+1D, the fermionic MCG is generated by $T^2$
and $S$.} Since the anomaly corresponds to a topological order in one higher
dimension that is described by a higher category, the change of anomalous
partition function can be described by the data of this higher category.  In
this paper, we will derive one such result.

Consider a CFT in $d$-dimensional closed space-time $M^d$, whose gravitational
anomaly is described a $(d+1)$D topological order.  The $(d+1)$D topological
order has $N$-fold degenerate ground states on $M^d$.  Let $G_{M^d}$ be the MCG
for $M^d$.  Under a MCG transformation $g \in G_{M^d}$, the degenerate ground
states transform according to a representation $R^\text{top}(g)$ of
$G_{M^d}$.\cite{W9039,KW9327,RSW0777,KW1458} Such a representation
$R^\text{top}(g)$ is the data that characterize the $(d+1)$D topological order
(and hence the anomaly).  It was conjectured\cite{W9039} that such data fully
characterize the topological order.  From the correspondence described in Fig.
\ref{BdryGrnd}, we find that \frmbox{the anomalous CFT in $d$-dimensional
space-time has several partition functions $Z(g_{\mu\nu},i)$,
$i=1,2,\cdots,\dim(R^\text{top})$, which transform as:
\begin{align}
\label{ZRZ}
Z(g\cdot g_{\mu\nu},i) &=
 R^\text{top}_{ij}(g) Z(g_{\mu\nu},j) ,
\end{align}
where $g_{\mu\nu}$ is the metrics on the  $d$-dimensional space-time $M^d$,
which describes the shape of $M^d$, and $g\cdot g_{\mu\nu}$ is the MCG action
on $g_{\mu\nu}$.} When $d=2$, \eqn{ZRZ} becomes \eqn{ZST}, which we will
explain in some detail.

For an anomaly-free CFT, the corresponding $(d+1)$D topological order is
trivial and $R^\text{top}=1$ are 1-by-1 matrices.  In the case, the above
becomes the usual MCG invariant condition on the partition function:
\begin{align}
 Z(g\cdot g_{\mu\nu})= Z(g_{\mu\nu}) .
\end{align}
It is likely that the MCG invariant partition functions on $M^d$ completely
classify anomaly-free CFTs.  Thus, it is also likely that that the modular
covariant partition function \eq{ZRZ} completely classify anomalous CFTs (\ie
the boundaries of $(d+1)$D topological order described by $R^\text{top}(g)$).

We like to point out that \eqn{ZRZ} also covers the cases of gapped boundaries
of $(d+1)$D topological order. In this case $Z(g_{\mu\nu},i)=Z(i)$ becomes
$g_{\mu\nu}$ independent.  The $d=2$ case is studied in detail in
\Ref{LWW1414,dwall}, where $Z(i)$ is denoted as $W^{\one i}$ and is called
fusion matrix or wavefunction overlap.  Thus \eqn{ZRZ} is a unified description
for both gapped and gapless boundary.

As another application, we point out that anomaly-free fermionic theories
exactly correspond a subset of the bosonic theories with the non-invertible
gravitational anomaly described by the bosonic $Z_2$ topological order with
emergent fermion (\ie the twisted $Z_2$ gauge theory) in one higher dimension.
Thus we can construct anomaly-free fermionic theories, such as their partition
functions, by constructing bosonic theories and their partition functions with
this particular non-invertible gravitational anomaly.

We want to mention that \Ref{KZ170501087} has given a very general and complete
theory of purely chiral CFT on the boundary of a 2+1D topological order, based
on tensor category theory.  The general theory developed here works for both
purely chiral and non-chiral  CFT on the boundary.  When the boundary is purely
chiral, our theory is just a subset of the full theory developed in
\Ref{KZ170501087}.  Both theories provide a unified approach for gapped and
gapless boundaries. Recently, some non-chiral CFT's on the boundary of 2+1D
topological order were studied in \Ref{CZ190312334}.  In particular, multi-component
partition functions on a 1+1D gapless boundary of 2+1D double-Ising topological
order were calculated.  A connection between the modular transformation of
boundary partition functions and the $S,T$ matrices that characterize the
modular tensor category for the 2+1D bulk topological order was noticed.  The
appearance of many sectors in anomalous CFT was also pointed out in
\Ref{LS190404833}.  Our paper generalizes those results and provides a more
systematic discussion.

We also like to remark that, in the presence of symmetry, there are also
several partition functions from the different symmetry twisting boundary
conditions in $d$-dimensional space-time.\cite{RZ1232,TR171004730}  If the
anomaly is not invertible, there will be several partition functions for each
twisted  boundary condition.  Those partition functions also transform
covariantly under MCG transformations.  This generalization is discussed in
\Ref{STanomaly}, for $d=2$ case.

\section{Topological invariant and properties of boundary partition function}

First, let us describe the topological path integral that can realize various
topological orders.  The boundaries of those topological orders realize
invertible and non-invertible anomalous theories.  This way, we can relate
anomalies with topological invariants in one higher dimensions. 

\subsection{Topological partition function as topological invariant}

A very general way to characterize a topologically ordered phase is via its
partition function $Z(M^D)$ on closed spactime $M^D$ with all possible
topologies. A detailed discussion on how to define the partition function via
tensor network is given in \Ref{KW1458} and in Appendix \ref{path}.  From
this careful definition, we see that the partition function also depends on the
branched triangulation of the space-time (see Appendix \ref{path}), as well
as the tensor associated with each simplex.  We collectively denote the
triangulation, the branching structure, and the tensors as $\cT$.  Thus the
partition function should be more precisely denoted as $ Z_\text{TN}(M^D,\cT)$.  
In a very fine triangulation limit (\ie the thermodynamic limit), we believe
that the partition function depends on $\cT$ via an effective metric tensor
$g_{\mu\nu}$ of the spacetime manifold, if the tensor network describes a
``liquid'' state, as opposed to a foliated state (a non-liquid
state).\cite{C0502,H11011962,ZW1490,SC171205892,SC180310426} Thus, the
partition function can be denoted as $Z_\text{field}(M^D,g_{\mu\nu})$ in the
thermodynamic limit.  $Z_\text{field}(M^D,g_{\mu\nu})$ correspond to the
partition function of a field theory where the different lattice
regularizations $\cT$ are not important as long as they produce the same
equivalent metric $g_{\mu\nu}$.  Here, $g_{\mu\nu}$ and $g'_{\mu\nu}$ are
regarded as equivalent if they differ by a diffeomorphim since
$Z_\text{field}(M^D,g_{\mu\nu}) = Z_\text{field}(M^D,g'_{\mu\nu})$.  Let
$\cM_{M^D}$ be the space formed by all metrics $g_{\mu\nu}$ of $M^D$ (up to
diffeomorphic equivalence), which is called the moduli space of $M^D$. Thus the
partition function $Z_\text{field}(M^D,g_{\mu\nu})$ is a complex function on
the moduli space $\cM_{M^D} \xrightarrow{Z_\text{field}(M^D, -)}  \C$.

However, $Z_\text{TN}(M^D,\cT)$ (or $Z_\text{TN}(M^D,g_{\mu\nu})$) is not a
topological invariant since it contains a so-called volume term
$\ee^{-\int_{M^D} \eps\, \dd^D x }$ where $\eps$ is the energy density.
But this problem can be fixed, by
factoring out the volume term.
This way, we can obtain a topological partition function
$Z_\text{TN}^\text{top}(M^D)$ which is believed to be a topological
invariant:\cite{KW1458,WW180109938}
\begin{align}
 Z_\text{TN}(M^D,\cT) = \ee^{-\int_{M^D} \eps\, \dd^D x } Z_\text{TN}^\text{top}(M^D,\cT) .
\end{align}
Appendix \ref{toppath} describes the way to fine-tune the tensors to make the
volume term vanishes (\ie $\eps=0$). In this case, the path integral directly
produces the topological partition function. 
Such a topological invariant may completely characterize the topological order.

Let us describe topological invariant, the topological partition function of the
field theory, $Z_\text{field}^\text{top}(M^D,g_{\mu\nu})$ (\ie
$Z_\text{TN}^\text{top}(M^D,\cT)$) in more details.
The ``topological property'' of $Z_\text{field}^\text{top}(M^D,g_{\mu\nu})$
 may appear in two ways:\cite{KW1458}
\begin{enumerate}
\item
 $Z_\text{field}^\text{top}(M^D,g_{\mu\nu})$ is a local constant function on
$\cM_{M^D}$. In this case, the topological partition function only depends on
$\pi_0(\cM_{M^D})$: $\pi_0(\cM_{M^D}) \xrightarrow{Z_\text{field}(M^D, -)}
\C$.  Such a complex function on $\pi_0(\cM_{M^D})$ is a topological invariant,
since $Z_\text{field}^\text{top}(M^D,g_{\mu\nu})$ does not depend on any smooth
change of $g_{\mu\nu}$.  In this case, the boundary has a global gravitational
anomaly. 

\item
 The reduction from lattice partition function
$Z_\text{TN}^\text{top}(M^D,\cT)$ to the field theory partition function
$Z_\text{field}^\text{top}(M^D,g_{\mu\nu})$ may have a phase ambiguity.
However, we can define the change of phase for
$Z_\text{field}^\text{top}(M^D,g_{\mu\nu})$ as we go along a segment $I$ in
$\cM_{M^D}$ without ambiguity: 
\begin{align} 
\text{phase change}= \ee^{\ii 2\pi \oint_{I} \al}= \ee^{\ii 2\pi \oint_{M^D\times I} \Om} 
\end{align} 
where $\al$ is a 1-form on $\cM_{M^D}$ and $\Om$ is closed $D+1$ form
constructed from the curvature tensor on $M^D\times I$. In this case, the
boundary has a perturbative gravitational anomaly.  For example, when $D=3$,
$\Om=\frac{\Del c}{24} p_1$, where $p_1$ is the first Pontryagin class on
4-manifold.  $Z_\text{field}^\text{top}(M^3,g_{\mu\nu})$ is given by
\begin{align}
 Z_\text{field}^\text{top}(M^3,g_{\mu\nu}) =
\ee^{\ii \frac{2\pi \Del c}{24} \oint_{M^D} \om_3} 
\end{align}
where the 3-form $\om_3$ satisfies $\dd \om_3 = p_1$ and corresponds to the
gravitational Chern-Simons term.  The coefficient $\Del c$ is the chiral
central charge of the boundary state.  In this case,
$Z_\text{field}^\text{top}(M^3,g_{\mu\nu})$ depends on the smooth change of
$g_{\mu\nu}$ and is not a topological invariant in the usual sense.

\end{enumerate}

\subsection{Invertible and non-invertible topological order}

Most topological orders are not invertible under the stacking operation.  (Here,
by definition, an invertible order\cite{KW1458,F1478} can be canceled by
another order, \ie the stacking of the two orders gives rise to the trivial
order.)  The invertible topological orders form a subset of topological orders.
The topological invariant for invertible topological orders,
$Z_\text{TN}^\text{top}(M^D,\cT)$, is a pure phase factor:
$|Z_\text{TN}^\text{top}(M^D,\cT)|=1$.\cite{W1447,HW1339,K1459,KW1458}

The boundaries of invertible topological orders have the standard gravitational
anomalies. The gravitational anomalies in literature all belong to this case.
The boundaries of non-invertible topological orders represent the structures
that are different from the usual anomalies.  We will call those structures as
non-invertible gravitational anomalies, and call the standard gravitational
anomalies as invertible gravitational anomalies.  In this paper, we will
concentrate on the non-invertible anomalies.

To give an example of invertible anomalies, let us consider a $E_8$ bosonic
quantum hall state described by the following $K$-matrix\cite{WZ9290,W9505}
\begin{align}
\label{KE8}
\nonumber\\
 K_{E_8} &=
{\footnotesize
\begin{pmatrix}
 2&-1& 0& 0& 0& 0& 0& 0\\
-1& 2&-1& 0& 0& 0& 0& 0\\
 0&-1& 2&-1& 0& 0& 0& 0\\
 0& 0&-1& 2&-1& 0& 0& 0\\
 0& 0& 0&-1& 2&-1& 0&-1\\
 0& 0& 0& 0&-1& 2&-1& 0\\
 0& 0& 0& 0& 0&-1& 2& 0\\
 0& 0& 0& 0&-1& 0& 0& 2\\
\end{pmatrix}
}
,
\end{align}
which has an invertible topological order, since det$(K_{E_8})=1$.  Its
boundary is described by the $(E_8)_1$ CFT that has a perturbative
gravitational anomaly, due to its non-zero chiral central charge $c=8$.  It is
a chiral CFT whose partition function has a single character,
\begin{align}
Z(\tau)=\chi^{E_8}(\tau)=\frac{\Theta_{K_{E_8}}(q)}{\eta^8 (q)},\ \ \
q\equiv \ee^{2\ii
\pi \tau}
\end{align}
where $\eta (q)=q^{\frac{1}{24}}\prod_{n=1}^{\infty} (1-q^n)$ is the Dedekind
eta function, and $\Theta_{K}$ is the theta function for a lattice
characterized by an integer symmetric matrix $K$:
\begin{align}
 \Theta_{K}(q) &= \sum_{\v n \in \Z^{\text{dim}K}} q^{\v n^\top K \v n/2}.
\end{align}
and $K_{E_8}$ is the $E_8$ root lattice, given by \eqn{KE8}.  The first a few
terms in the expansion is
\begin{align}
\chi^{E_8}=q^{-1/3}(1+248q+4124q^2+O(q^3)),
\end{align} 
where the $248$ generators of $E_8$ are counted in the second term in this single
sector.  $\chi^{E_8}$ transforms according to the one-dimensional
representation of the modular group
\begin{align}
\chi^{E_8}(-1/\tau)=\chi^{E_8}(\tau),\quad \chi^{E_8}(\tau+1)=
\ee^{-\ii \frac{2\pi}{3}}\chi^{E_8}(\tau)
\end{align}

The 1+1D perturbative gravitational anomaly characterized by the chiral central
charge $\Del c$ constrains the boundary partition function in 1+1D:
\begin{align}
\label{pga}
\lim_{q\to 0} 
Z(q)  = \text{integer}\times q^{-\frac{c}{24}} \bar q^{-\frac{\bar c}{24}}, \ \ \ \ \ \
\Del c=c-\bar c.
\end{align}
Thus, knowing the 1+1D boundary partition function, we can also determine its
perturbative gravitational anomaly $\Del c$.  In this paper, we will try to go
one step further. We like to determine the global anomaly from the
partition function.  

\subsection{Properties of boundary partition function}

To concentrate on global anomaly, we will assume that there is no perturbative
anomaly.  In this case, the global anomaly is characterized by the bulk
topological invariant $Z_\text{field}^\text{top}(M^D,\cT)$, which  can be
realized by the topological path integral described in Appendix
\ref{toppath}.\cite{KW1458} In this paper, we assume the bulk theory is always
described by the topological path integral, whose partition function directly
corresponds to the topological invariant $Z_\text{field}^\text{top}(M^D,\cT)$.

%One way to systematically generate some of the bulk topological invariant
%$Z_\text{field}^\text{top}(M^D,A,\cT)$ is to consider the following topological
%partition function $Z^\text{top}(B^d \gext_\vphi S^1,A)$.  Here where $d=D-1$
%and $B^d \gext_\vphi S^1$ is the mapping torus obtained from $I\times B^d$ by
%gluing its two boundaries via map $\vphi$: $B^d\to B^d$ in the mapping class
%group of $B^d$.  In other words, $B^d \gext_\vphi S^1$ is a fiber bundle with
%fiber $B^d$ and base space $S^1$.  We see that, we can obtain a topological
%invariant from each element of mapping class group of $B^d$, provided that
%there is no perturbative anomaly.

\begin{figure}[tb]
\begin{center}
\includegraphics[scale=0.55]{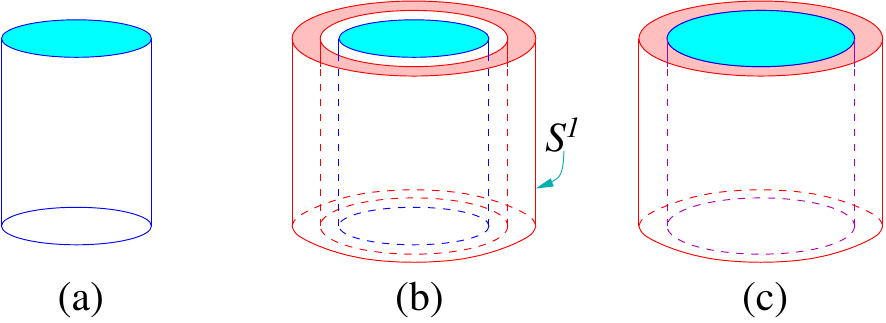} 
\end{center}
%Fig. 1
\caption{ 
(a) Space-time $D^2\times S^1$ (solid cylinder). (b) $I\times S^1\times S^1$
(cylinder) and $D^2\times S^1$ (solid cylinder). (c) Gluing the cylinder with
solid cylinder, along the $S^1\times S^1=T^2$ boundary, reproduces the
space-time  $D^2\times S^1$.  The tensor networks on the solid cylinder and the
cylinder define the path integral.  The tensors on the inner solid cylinder are
the bulk tensors that describe a topological path integral.  The tensors on the
outer cylinder can be anything, which may describe a gapless CFT at long
distance.  Different choices of boundary tensor network on the outer cylinder
give rise to different types of boundaries.
}
\label{bdyblk}
\end{figure}

To link such a topological invariant (\ie topological path integral),
$Z_\text{TN}^\text{top}(M^D,\cT)$ to the partition function on the boundary
$B^d$, $d=D-1$, we note that the  boundary partition function is given by (Fig.
\ref{bdyblk}a)\cite{KW1458,WW180109938}
\begin{align}
 Z(B^d; M^D,\cT) =  Z_\text{TN}^\text{top}(M^D,\cT),\ \ \ \
B^d=\prt M^D.
\end{align}
The boundary is the so-called natural boundary described in Appendix
\ref{Nboundary}, but here we sum over the boundary degrees of freedom.  We note
that the bulk is gapped. Thus, the low energy properties of the boundary (below
the bulk gap) are described by the above $Z(B^d,\cT_B)$.

We may obtain a more general boundary by stacking a $d$-dimensional system
described by a $d$-dimensional tensor network, $Z_\text{TN}(B^d,\cT_B)$, to
the boundary (see Fig. \ref{bdyblk}b). The resulting boundary partition function has a form
\begin{align}
 Z(B^d,\cT_B; M^D, \cT) =  Z_\text{TN}(B^d,\cT_B) Z_\text{TN}^\text{top}(M^D,\cT)
\end{align}
We may also allow the boundary and bulk degrees of freedom to interact with
each other by gluing the boundary to the bulk as in Fig. \ref{bdyblk}c. We
see that the boundary partition function $Z(B^d,\cT_B;M^D,\cT)$ is not purely
given by a tensor network on the boundary $B^d$, which gives rise to a
partition function $Z_\text{TN}(B^d,\cT_B)$.  The boundary partition function
also contain a bulk topological term $Z_\text{TN}^\text{top}(M^D,\cT)$. This
makes the boundary quantum system defined by $Z(B^d,\cT_B;M^D,\cT)$ to be
potentially anomalous.  If the boundary partition function is given purely by a
tensor network $Z_\text{TN}(B^d,\cT_B)$ on the boundary (\ie when
$Z_\text{TN}^\text{top}(M^D,\cT)=1$), such a quantum system will be
anomaly-free.

\subsection{1+1D anomalous theory on space-time torus $T^2$}

In this section, we will concentrate on an 1+1D anomalous theory.  To define
its partition function on a  space-time torus $T^2$, we consider a 2+1D tensor
network path integral (see Appendix \ref{path}) on $D^2\times S^1$
(see Fig. \ref{bdyblk}c)
\begin{align}
\label{Zboundary}
 Z(T^2; D^2\times S^1) =  Z_\text{TN}^\text{top}(D^2\times S^1), \ \ \ \
\prt D^2 = S^1.
\end{align}
The tensors on the inner solid cylinder define a topological path integral
described in Appendix \ref{path} which realize a topological order that
corresponds to the anomaly under consideration.  The tensors on the outer
cylinder (see Fig. \ref{bdyblk}b) can be anything, which determine the
different types of boundaries.

\begin{figure}[tb]
\begin{center}
\includegraphics[scale=0.55]{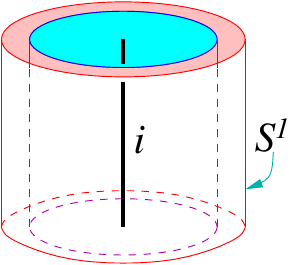} 
\end{center}
%Fig. 1
\caption{ 
The space-time $D^2\times S^1$ with a world-line of type-$i$ topological
excitation, wrapping in the $S^1$ direction.  The path integral on the inner
solid cylinder is a topological path integral with world-line, as described in
Appendix \ref{toppathW}.
}
\label{bdyblkW}
\end{figure}

We can define a more general partition function for 1+1D anomalous theory by
inserting a world-line (see Fig. \ref{bdyblkW})
\begin{align}
 Z(T^2; D^2_i\times S^1) =  Z_\text{TN}^\text{top}(D^2_i\times S^1).
\end{align}
We note that the surface of the inner solid cylinder in  Fig. \ref{bdyblkW}
(after integrating out only the bulk degrees of freedom as in Appendix
\ref{Nboundary}) corresponds to a wavefunction, $|\psi_i\>$, that describes one
of the degenerate ground states of the bulk topological order on the torus. If
the path integral on the inner solid cylinder is a topological path integral,
$|\psi_i\>$ automatically normalizes to 1: $\<\psi_i|\psi_i\>=1$ (as discussed
in \Ref{dwall}).  Thus, more precisely, the 1+1D partition function for an
anomalous theory is given by
\begin{align}
\label{Zboundary}
 Z(T^2, |\psi_i\>) =  Z_\text{TN}^\text{top}(D^2_i\times S^1).
\end{align}
Here the degenerate ground state wave functions $|\psi_i\>$ are labeled by the
type-$i$ of the topological excitations.  For the trivial excitations labeled
by $\one$, $Z(T^2, |\psi_\one\>)$ correspond to the partition function for the
space-time in Fig. \ref{bdyblk}c without the insertion of the world-line.

The dependence on the ground state wave function $|\psi_i\>$ of the topological
order on the torus is the key character of anomalous partition function:
\begin{enumerate}
\item
If $|\psi_i\>$ is a product state, then
$Z(T^2, |\psi_i\>)$ is a partition function of an anomaly-free theory.

\item
If $|\psi_i\>$ is unique (\ie the topological order has a non-degenerate ground
state on the torus), then $Z(T^2, |\psi_i\>)$ is a partition function of a theory
with invertible anomaly.

\item
If $|\psi_i\>$ is not unique (\ie the topological order has degenerate ground
states on torus), then $Z(T^2, |\psi_i\>)$ is a partition function of a theory
with non-invertible anomaly.

\end{enumerate}

\subsection{Modular transformations of the partition function for an anomalous
theory }

Let us fine tune the action of the 1+1D anomalous theory, so that it  has a
vanishing ground state energy density.  In this case, its partition function on
$T^2$ will not depend on the size of the space-time, but only depend on the
shape of the space-time.  The shape of a torus $T^2$ can described by a complex
number $\tau$. Thus we may write the 1+1D partition function as  
\begin{align}
\label{Zb}
 Z(\tau,\bar\tau, |\psi_i\>) =  Z_\text{TN}^\text{top}(D^2_i\times S^1).
\end{align}
However, $\tau$ and $\tau'=\tau+1$ describe the same shape after a coordinate
transformation.  For an anomaly-free 1+1D theory, we have
\begin{align}
 Z(\tau,\bar\tau)=Z(\tau+1,\bar\tau+1).
\end{align}
However, for an anomalous 1+1D theory, we have
\begin{align}
\label{ZbT}
 Z(\tau,\bar\tau, T^\text{top}_{ij}|\psi_j\>) =  Z(\tau+1,\bar\tau+1, |\psi_i\>),
\end{align}
since the coordinate transformation acts non-trivially on the ground state
wavefunction $|\psi_i\>$ on torus.  Here the unitary matrix $T^\text{top}_{ij}$
describes such a non-trivial action, which is a modular transformation of the
torus ground states of the 2+1D bulk topological order.\cite{W9039,KW9327}
Similarly, $\tau$ and $\tau'=-1/\tau$ also describe the same shape after a
coordinate transformation.  Thus
\begin{align}
\label{ZbS}
 Z(\tau,\bar\tau, S^\text{top}_{ij}|\psi_j\>) =  Z(-1/\tau,-1/\bar\tau, |\psi_i\>),
\end{align}
where the unitary matrix $S^\text{top}_{ij}$ describes another modular
transformation of the torus ground states of the bulk topological order.

We note that the partition function $Z(\tau,\bar\tau, |\psi_i\>)$ depends on $|\psi_i\>$
in a linear fashion 
\begin{align}
Z(\tau,\bar\tau, M_{ij}|\psi_j\>) =  M_{ij} Z(\tau,\bar\tau, |\psi_j\>)
\end{align}
This is because the path integral that sums over the degrees of freedom in the
bulk and the outer surface of outer cylinder (see Fig. \ref{bdyblk}b) gives
rise to a wave function $\< \phi|$ that lives on the inner surface of the outer
cylinder.  The partition function $Z(\tau,\bar\tau, |\psi_i\>)$ is simply
\begin{align}
 \< \phi| \psi_i\> = Z(\tau,\bar\tau, |\psi_i\>) .
\end{align}
Thus $Z(\tau,\bar\tau, |\psi_i\>)$ is a linear function of $|\psi_i\>$.
As a result, \eqn{ZbT} and \eqn{ZbS} can be rewritten as
\begin{align}
\label{ZST}
 T^\text{top}_{ij} Z(\tau,\bar \tau; j) =&   Z(\tau+1,\bar\tau+1; i), \nn\\
 S^\text{top}_{ij} Z(\tau,\bar \tau; j) =&   Z(-1/\tau,-1/\bar\tau; i),
\end{align}
where $Z(\tau,\bar\tau, i) \equiv  Z(\tau,\bar\tau, |\psi_i\>)$.  \Eqn{ZST} is
another key result of this paper. It describes the modular transformation
properties of the partition functions for anomalous theory.  

We stress that, for an anomalous theory, its partition functions are vectors in
a vector space $\cZ$.  The anomalous theory itself only determines such a vector
space $\cZ$.  When $\cZ$ is one dimensional, the anomaly is invertible.  When
the dimension of $\cZ$ is more than one, the anomaly is non-invertible.

For gapped anomalous theory, the partition functions do not depend on $\tau$.
\Eqn{ZST} becomes
\begin{align}
 \label{ZST0}
 Z(i) =  T^\text{top}_{ij} Z(j), \ \ \ \ 
 Z(i) =  S^\text{top}_{ij} Z(j).
\end{align}
We recover a condition for gapped boundary of a topological order obtained in
\Ref{LWW1414,dwall}, where $Z(i)$ was denoted as $W^{\one i}$.  Note that for
the gapped case, the partition functions $Z(i)$ are ground state degeneracy of
the systems and are non-negative integers.

The above is a general discussion of 1+1D anomalous theory, which can have a
non-invertible anomaly. In particular, the boundary CFT may have separate
right-moving part and left-moving part, and each part transforms according to certain
$S_{R,L}$ and $T_{R,L}$ matrices.  Those boundary $S_{R,L}$ and $T_{R,L}$
matrices may be \emph{different} from the $S^\text{top}$ and $T^\text{top}$ matrices
for the bulk topological order.  However, after we combine the right movers and
left movers to construct multi component partition functions $ Z(\tau,\bar
\tau; i)$, we find that $ Z(\tau,\bar \tau; i)$ transform according to the bulk
$S^\text{top}$ and $T^\text{top}$ matrices. In the following, we will discuss
some simple examples of 1+1D non-invertible anomaly.

\section{A non-invertible bosonic global gravitational
anomaly from 2+1D $Z_2$ topological order}

A 2+1D $Z_2$ topological order has four type of excitations, $\one,e,m,f$,
where $e,m$ are bosons and $f$ is a fermion.  $e,m,f$ are topological
excitations with $\pi$ mutual statistics respect to each other.  (Remember that
a topological excitation is defined as the excitation that cannot be created by
any local operators).  Such a topological order can have many different
boundaries, which all carry the same non-invertible gravitational anomaly.  In
this section, we will discuss some of those boundary theories.\cite{KW1458}

\subsection{Two gapped boundaries of the 2+1D $Z_2$ topological order }
\label{gapped}

A gapped boundary of the 2+1D $Z_2$ topological order is induced by $m$
particle condensation. This boundary has only one type of topological
excitations $e$.  The topological excitation $e$ has a $Z_2$ fusion $e\otimes
e=1$, and is described by a symmetric fusion category $\Rep(Z_2)$ (which is the
fusion category formed by the representations of $Z_2$ group).  Such a boundary
described by $\Rep(Z_2)$ has a non-degenerate ground state.  Its partition
function is given by $Z(\tau,\bar\tau,\one)=1$ (where $\one$ means that there
is no insertion of world-line, \ie $i=\one$ in Fig. \ref{bdyblkW}).

The insertion of a world-line of $m$-type topological excitations (see Fig.
\ref{bdyblkW}) produces another boundary, where $e$ on the boundary $S^1$
acquires a $\pi$-phase as it goes around the boundary.  The partition function for
such a boundary is still given by $Z(\tau,\bar\tau,m)=1$.

If we insert a world-line of $e$-type or a $f$-type, the resulting boundary
will carry an un-paired $e$ excitations.  Such an un-paired $e$ costs a finite
energy $\eps_e$.  These boundaries will have partition functions
$Z(\tau,\bar\tau,e)=Z(\tau,\bar\tau,f)= \# \ee^{-\eps_e \bt } \big|_{\bt \to \infty} = 0$, when
the size of space-time $\bt$ approaches to infinity.

So the first gapped boundary of $Z_2$ topological order is described by four
partition functions in the excitations basis $(\one,e,m,f)$
\begin{align}
\label{bm}
Z(\tau,\bar\tau,\one)= Z(\tau,\bar\tau,m)=1,\ \ \  
Z(\tau,\bar\tau,e)= Z(\tau,\bar\tau,f)=0.
\end{align}
They can be viewed as the partition function for an anomalous $c=0$ CFT (\ie a
gapped theory).  One can check that these four partition functions in the
excitations basis satisfy \eqn{ZST0},\cite{LWW1414,dwall}
since for $Z_2$ topological order, $S^\text{top},T^\text{top}$ are given by
\begin{align}
\label{Z2STmat}
  T^\text{top}_{Z_2}&=
\begin{pmatrix}
    1&0&0&0\\
    0&1&0&0\\
    0&0&1&0\\
    0&0&0&-1
  \end{pmatrix} 
&
  S^\text{top}_{Z_2}&=\frac12 \begin{pmatrix}
    1&1&1&1\\
    1&1&-1&-1\\
    1&-1&1&-1\\
    1&-1&-1&1
  \end{pmatrix}
\end{align}

Let us obtain another gapped boundary of the 2+1D $Z_2$ topological order, by
lowering the energy of $e$ to a negative value. This will drive a ``$Z_2$
symmetry'' breaking transition and obtain an $e$-condensed state, which have a
2-fold ground state degeneracy on a ring.  (If we condensed $e$ particle on an
open segment on the boundary, we will also get a 2-fold ground state
degeneracy.) This new boundary is described by the following four partition
functions 
\begin{align}
\label{be2}
Z(\tau,\bar\tau,\one)= Z(\tau,\bar\tau,e)=2,\ \ \ \ 
Z(\tau,\bar\tau,m)= Z(\tau,\bar\tau,f)=0.
\end{align}
They again satisfy \eqn{ZST}.

Here $Z(\tau,\bar\tau,\one)=2$ means the $Z_2$ topological order on $D^2$ (\ie
the boundary state on $S^1$, see Fig. \ref{bdyblk}c) has a 2-fold degeneracy.
This 2-fold degeneracy come form the emergent mod-2 conservation of
$e$-particles on the boundary, and subsequently the spontaneous breaking of
this emergent $Z_2$ symmetry.  However, since the $Z_2$ symmetry is emergent,
when the boundary $S^1$ has a finite density $n_e$ of the $e$-particles, the
emergent mod-2 conservation may be explicitly broken by an amount $\ee^{-1/n_e
\xi}$ where $\xi$ is a length scale.  In this case, the   2-fold degeneracy is
lifted by an amount $\ee^{-1/n_e \xi}$.  So the boundary described by \eqn{be2}
is unstable.  After the lifting of the degeneracy, the boundary is actually
described by
\begin{align}
\label{be}
Z(\tau,\bar\tau,\one)= Z(\tau,\bar\tau,e)=1,\ \ \
Z(\tau,\bar\tau,m)= Z(\tau,\bar\tau,f)=0,
\end{align}
which correspond to the boundary of the 2+1D $Z_2$ topological order induced by
$e$ condensation (while the boundary induced by $m$ condensation is
described by \eqn{bm}).

\subsection{A gapless boundary of the 2+1D $Z_2$ topological order }
\label{gapless}

A gapless boundary of the 2+1D $Z_2$ topological order is given by a 1+1D
gapless system described by a Majorana fermion field 
\begin{align}
 H = \int \dd x\ \la_R \ii \prt_x \la_R  -\la_L \ii \prt_x \la_L .
\end{align}
We like to stress that such a  1+1D gapless system is actually a bosonic system
where \emph{the states in the many-body Hilbert are all bosonic (\ie contain an
even number of Majorana fermions)}.  We refer such a 1+1D gapless system as the
boson-restricted  Majorana fermion theory.  It is different from the usual Majorana
fermion theory.

We can give the Majorana fermion a mass gap to obtain a gapped boundary:
\begin{align}
 H = \int \dd x\ \la_R \ii \prt_x \la_R  -\la_L \ii \prt_x \la_L + \ii m \la_R\la_L
.
\end{align}
Such a gapped boundary correspond to the gapped boundary described above.  If
we lower $m$ to a negative value, we should drive the ``$Z_2$ symmetry''
breaking transition described above and obtain a 2-fold ground state degeneracy
on a ring.  This is different from the standard Majorana fermion theory where
the negative $m$ also gives rise to non-degenerate ground state.  So for our
boson-restricted Majorana fermion theory, a positive $m$ gives rise to
non-degenerate ground state while a negative $m$ gives rise to a 2-fold ground
state degeneracy on a ring.  If we only change the sign of $m$ on an open
segment, then both the standard Majorana fermion theory and our bosonic
Majorana fermion theory will give rise to  a 2-fold ground state degeneracy.  

So when $m=0$ the gapless bosonic  Majorana fermion theory describes the
critical point of the $Z_2$ symmetry breaking phase transition mentioned above.
The gapless boson-restricted Majorana fermion theory describes a conformal field
theory (CFT) with a non-invertible gravitational anomaly.  In this paper, we
like to understand this anomalous CFT in detail. In particular, we would like
to compute its partition function and their properties under modular
transformation.

To understand the critical CFT for the ``$Z_2$ symmetry'' breaking transition,
let us introduce a 1d lattice Hamiltonian \emph{on a ring} to describe the
gapped boundary in Section \ref{gapped}
\begin{align}
\label{tIsing}
H = - U \sum_i  \si^z_i - J \sum_i \si^x_i \si^x_{i+1}, \ \ \ U,J >0
\end{align}
where $\si^l,\ l=x,y,z$  are Pauli matrices.  Here an up-spin $\si^z_i=1$
correspond to an empty site and an down-spin $\si^z_i=-1$ correspond to a site
occupied with an $e$ particle.  Since number of the $e$ particles is always
even, thus \emph{the Hilbert space $\cV$ of our model is formed by states with
even numbers of down spins $ \si^z_i=-1$.} Note that our Hilbert space is
\emph{non-local}, \ie it does not have a tensor product decomposition:
\begin{align}
\cV \neq \otimes_i \cV_i 
\end{align}
where $\cV_i$ is the two dimensional Hilbert space for site-$i$.  It is this
property that make our model to have a non-invertible gravitational anomaly.

We like to mention that, we can view the 2+1D $Z_2$ topological order as a
gauged $Z_2$ symmetric state with a trivial SPT order.  The boundary of the
2+1D $Z_2$ symmetric state can be described by a transverse Ising model
\eqn{tIsing} with the standard Hilbert space (\ie without the $\prod_i
\si^z_i=1$ constraint).  The boundary can be in a symmetric phase (described by
\eqn{tIsing} with $U\gg J$) or a $Z_2$ symmetry breaking phase (described by
\eqn{tIsing} with $U\ll J$).  We see that after gauging the $Z_2$ symmetry to
obtain the $Z_2$ topological order in the bulk, the only change in the boundary
theory is the addition of the constraint $\prod_i \si^z_i=1$,\cite{YS13094596}
that changes the many-body Hilbert space to make it non-local (\ie make the
boundary theory to have a non-invertible gravitational anomaly).

In our model \eq{tIsing} for the boundary, the $J$ term is allowed since the
$e$ particles have only a mod-2 conservation.  In the $U \gg J$ limit, the
above lattice model describes the gapped phase in Section \ref{gapped}. As we
change $U$ to $U \ll J$, we will drive a ``$Z_2$ symmetry'' breaking phase
transition.  The critical point at $U=J$ is described by a CFT with
non-invertible gravitational anomaly.  Such a CFT is described by the
boson-restricted Majorana fermion theory mentioned in Section \ref{gapless}.
The Majorana fermion theory is obtained from \eqn{tIsing} via the Jordan-Wigner
transformation.

To obtain the partition function of the anomalous CFT, let us first consider
the partition function of the transverse Ising model \eq{tIsing} at critical
point $U=J$.  There are four partition functions $Z_{a_x,a_t}$ for the
transverse Ising model, with different $\Z_2$ boundary conditions $a_x=\pm1$
and $a_t=\pm1$.  The partition functions are given by the characters $
\chi_\one(\tau), \chi_\psi(\tau), \chi_\si(\tau), $ of two Ising CFTs (see
Appendix \ref{minCFT}), one for right movers and the other for the left movers.
We find
\begin{align}
Z_{1,1}=& |\chi_\one|^2+|\chi_\psi|^2+|\chi_{\si}|^2\nn\\
 Z_{1,-1}=&|\chi_\one|^2+|\chi_\psi|^2-|\chi_{\si}|^2\nn\\
Z_{-1,1}=& \chi_\one\bar\chi_{\psi}+\chi_\psi\bar\chi_\one+|\chi_{\si}|^2\nn\\
 Z_{-1,-1}=&-\chi_\one\bar\chi_{\psi}-\chi_\psi\bar\chi_\one+|\chi_{\si}|^2
\end{align}
This means that
the partition functions for the even and odd $Z_2$ sectors are  given by
\begin{align}
 Z_\text{even} &= \frac{Z_{1,1}+Z_{1,-1}}{2}=|\chi_\one|^2+|\chi_\psi|^2,
\nonumber\\
 Z_\text{odd} &= \frac{Z_{1,1}-Z_{1,-1}}{2}= |\chi_{\si}|^2.
\end{align}

For the anomalous CFT on the boundary of 2+1D $Z_2$ topological order, its
partition function is given by the  partition function of the Ising model for
the even $Z_2$ sector
\begin{align}
 Z(\tau,\bar\tau,\one)=|\chi_\one|^2+|\chi_\psi|^2
\end{align}
If we insert the $e$ world-line in the bulk (see Fig. \ref{bdyblk}),
the corresponding partition function $Z(\tau,\bar\tau, e)$ is given by $Z_\text{odd}(\tau,\bar\tau)$:
\begin{align}
 Z(\tau,\bar\tau, e) & = |\chi_\si|^2 
\end{align}
Similarly, we find
\begin{align}
 Z(\tau,\bar\tau, m) & = |\chi_\si|^2 
\end{align}
and
\begin{align}
 Z(\tau,\bar\tau, f) & = \chi_\one\bar\chi_{\psi}+\chi_\psi\bar\chi_\one 
\end{align}
We find that the above partition functions $Z(\tau,\bar\tau,i)$, $i=\one,e,m,f$, indeed
satisfy \eqn{ZST}.  Those partition functions describe a 1+1D gapless theoy
with a non-invertible gravitational anomaly, which can appear as a boundary of
the 2+1D $Z_2$ topological order.

\section{A non-invertible bosonic global gravitational
anomaly from 2+1D DS topological order}

Now let us consider the boundary of the 2+1D DS topological order.
Since the DS topological order can be viewed as a gauged 2+1D $Z_2$
symmetric state with the non-trivial $Z_2$ SPT order, we will first consider
the boundary theory of the 2+1D $Z_2$ SPT state on a 1d ring with even number
of sites:\cite{CW1235}
\begin{align}
\label{CZXb}
H &= - U \sum_i  \si^z_i\si^z_{i+1} - J \sum_i (\si^x_i + \si^z_{i-1}\si^x_i \si^z_{i+1}), 
\nonumber\\
&  U,J >0
\end{align}
The above Hamiltonian has a non-on-site $Z_2$ symmetry generated by
\begin{align}
\label{XCZ}
 U=\prod_i \si^x_i \prod_i CZ_{i,i+1}
\end{align}
where $CZ_{ij}$ acts on two spins as 
\begin{align}
\label{CZij}
CZ_{ij}&= |\up\up\>\<\up\up| +|\down\up\>\<\down\up|
+|\up\down\>\<\up\down| -|\down\down\>\<\down\down| .
\nonumber\\
&= \frac {1 +\si^z_i +\si^z_j -\si^z_i \si^z_j}2
\end{align}
From Appendix \ref{Z2trns}, we see that the above Hamiltonian in \eqn{CZXb} is
$Z_2$ symmetric. But the $Z_2$ symmetry has a 't Hooft anomaly.

To have a theory that is defined on rings with both even and odd sites, we
should consider different (but equivalent) non-on-site $Z_2$ symmetry:
\begin{align}
\label{Xs}
 U=\prod_i \si^x_i \prod_i s_{i,i+1}
\end{align}
where $s_{ij}$ acts on two spins as 
\begin{align}
\label{sij}
s_{ij}&= |\up\up\>\<\up\up| +|\down\up\>\<\down\up|
-|\up\down\>\<\up\down| +|\down\down\>\<\down\down| .
\nonumber\\
&= \frac 12( 1 -\si^z_i +\si^z_j +\si^z_i \si^z_j)
\end{align}
The $Z_2$ transformation has a simple picture: it flips all the spins and
include a $(-)^{N_{\up\to\down}}$ phase, where $N_{\up\to\down}$ is the number
of $\up \to \down$ domain wall .  From Appendix \ref{Z2trns}, we see that the
Hamiltonian \eqn{CZXb} is also invariant under the new $Z_2$ transformation.

The boundary of 2+1D $Z_2$ SPT state described by \eqn{CZXb} has a symmetry
breaking phase when $U\gg J$.  The boundary can also be gapless described by a
$c=\bar c=1$ CFT when $U=0$.  \Eqn{CZXb} has no symmetric gapped phase, since
the $Z_2$ symmetry is not on-site (\ie has a 't Hooft anomaly).\cite{CLW1141}

\begin{figure}[tb]
\begin{center}
\includegraphics[scale=0.4]{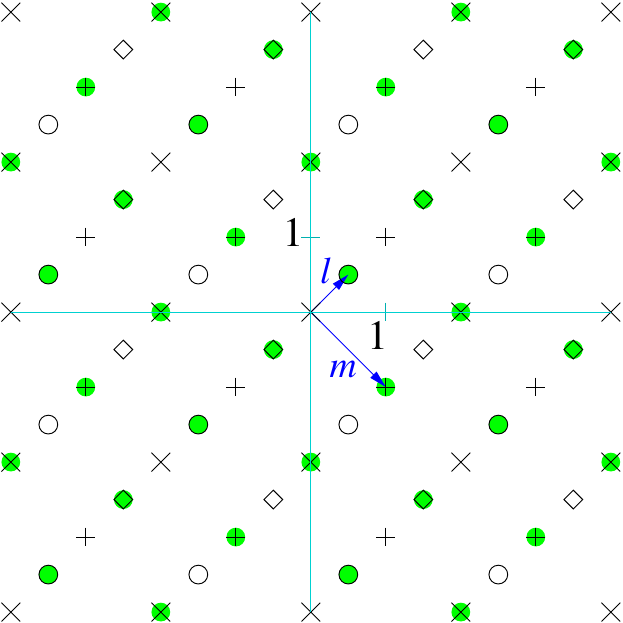} \end{center}
%Fig. 1
\caption{ 
The lattice $\Ga$ formed by points $(a,b)$.  Each point corresponds to a $U(1)$
vertex operator with scaling dimension $(h,\bar h)=(\frac 12 a^2,\frac 12
b^2)$.  The ``$\times$'' points give rise to $|\chi_0^{u1_4}|^2$.  The
``$\circ$'' points give rise to $|\chi_\one^{u1_4}|^2$.  The ``$+$'' points give
rise to $|\chi_2^{u1_4}|^2$.  The ``$\diamond$'' points give rise to
$|\chi_3^{u1_4}|^2$.  We also mark the directions of the $l$-label and
$m$-label.  The shaded points carry the $Z_2$-charge $l+m= 1$ mod 2, and the
unshaded points carry the $Z_2$-charge $l+m=0$ mod 2.} 
\label{U1latt} 
\end{figure}

When $U=0$, the model \eq{CZXb} can be mapped to the XY-model on 1d
lattice:\cite{CW1235} 
\begin{align}
 H_\text{XY} = -J \sum_i 
(\si^x_i \si^x_{i+1} + \si^y_i \si^y_{i+1}) .
\end{align}
In this case the anomalous 1+1D theory is gapless and is described by a
$u1_4\times \overline{u1}_4$ CFT (see Appendix \ref{U1CFT}). It has a partition
function 
\begin{align}
\label{ZXY}
 Z_\text{XY}(q,\bar q) 
\equiv Z_\text{$Z_2$-SPT}(q,\bar q) 
= \sum_{i=0}^3 |\chi_i^{u1_4}(q)|^2
\end{align}
with primary fields of dimension $(h_i,\bar h_i) = (\frac {i^2}{8}, \frac
{i^2}{8})$.  The above partition can be rewritten as
\begin{align}
 Z_\text{$Z_2$-SPT}(q,\bar q) =\frac{1}{|\eta(q)|^2} \sum_{(a,b)\in \Ga} q^{\frac12 a^2} \bar q^{\frac12 b^2}
\end{align}
where $(a,b)$ form a lattice $\Ga$ (see Fig. \ref{U1latt}):
\begin{align}
 (a,b) = \frac12 (l+2m,l-2m), \ \ \ \ l,m \in \Z.
\end{align}
So all the $U(1)$ vertex operators in our XY-model
can be labeled by $(l,m)$
which have scaling dimension
\begin{align}
(h,\bar h)
= \left(\frac12 a^2, \frac12 b^2\right) 
= \left(\frac{(l+2m)^2}{8}, \frac {(l-2m)^2}{8} \right). 
\end{align}
This is the labeling scheme used in \Ref{CW1235}.
It was found that
the $U(1)$ vertex operator labeled by $(l,m)$ carry the $Z_2$-charge
$l+m$ mod 2.

We see that each character $|\chi_i^{u1_4}|^2$ contains $U(1)$ vertex operators
with different $Z_2$-charges.  Thus it is more convenient to rewrite the
partition function in terms of the $u1_{16}$ characters
\begin{align}
 Z_\text{$Z_2$-SPT}(q) &= 
\sum_{i=0}^7 |\chi_{2i}^{u1_{16}}|^2
+\sum_{i=0}^7 \bar\chi_{2i+4}^{u1_{16}} \chi_{2i}^{u1_{16}}
\end{align}
The $U(1)$ vertex operators in $|\chi_{2i}^{u1_{16}}|^2$ carry the $Z_2$-charge
$i$ mod 2.  The $U(1)$ vertex operators in $\bar
\chi_{2i+4}^{u1_{16}}\chi_{2i}^{u1_{16}}$ carry the $Z_2$-charge $i+1$ mod 2.

In the presence of the $Z_2$-symmetry, we can define 4-partition functions for
different $Z_2$-symmetry twists in the space and time directions
$(a_x,a_t)=(\pm1,\pm1)$.  $Z_\text{$Z_2$-SPT}$ is the partition function with
no symmetry twist $(a_x,a_t)=(1,1)$:
\begin{align}
 Z_{1,1} = \sum_{i=0}^7 |\chi_{2i}^{u1_{16}}|^2
+\sum_{i=0}^7 \chi_{2i+8}^{u1_{16}} \bar\chi_{2i}^{u1_{16}}.
\end{align}
with a $Z_2$-symmetry twist in time directions the terms with $Z_2$-charge 1
acquire a $-$ sign:
\begin{align}
  Z_{1,-1} = \sum_{i=0}^7 (-)^i|\chi_{2i}^{u1_{16}}|^2
-\sum_{i=0}^7 (-)^i\chi_{2i+8}^{u1_{16}} \bar\chi_{2i}^{u1_{16}}. 
\end{align}
After an $S$-transformation of $u1_{16}$ (see Apendix \ref{U1CFT}), we get
\begin{align}
 Z_{-1,1} = 
\sum_{i=0}^7 \chi_{2i+1}^{u1_{16}} \bar\chi_{2i+5}^{u1_{16}} 
+\sum_{i=0}^7 \chi_{2i+1}^{u1_{16}} \bar\chi_{2i+13}^{u1_{16}}. 
\end{align}
From $Z_{-1,1}$ we find
\begin{align}
  Z_{-1,-1} = 
\sum_{i=0}^7 (-)^i\chi_{2i+1}^{u1_{16}} \bar\chi_{2i+5}^{u1_{16}} 
-\sum_{i=0}^7 (-)^i\chi_{2i+1}^{u1_{16}} \bar\chi_{2i+13}^{u1_{16}}.
\end{align}
by adding a $-$ sign to the terms with $Z_2$-charge 1.

Now we gauge the $Z_2$ on-site symmetry in the 2+1D SPT state to obtain the
2+1D DS topological order.  The 2+1D DS topological order
has a gapped boundary which contains topological excitation $s$ that satisfies a
$Z_2$ fusion role $s\otimes s=1$.  The 1d particles with $Z_2$ fusion role are
described by one of the two fusion categories.  The first one is $\Rep(Z_2)$
mentioned in the last section.  The second one is a different fusion category,
which we refer as the semion fusion category.\cite{FNS0428,LW0510} Such a
gapped boundary can be described by \eqn{CZXb} in $U\gg J$ limit (\ie in the
$Z_2$ symmetry breaking phase), where the $Z_2$ domain walls correspond to the
boundary particle $s$.  The fusion of those domain walls is described by the
semion fusion category, provided that the fusion processes preserve the
non-on-site $Z_2$ symmetry \eq{Xs}, 

However, there is one problem with the above picture: in the $Z_2$ symmetry
breaking phase, all the domain wall configurations have 2-fold degeneracy
induced by the $Z_2$ transformation \eqn{Xs}.  To fix this problem, we need to
modify the many-body Hilbert space on a ring by imposing the constraint
\begin{align}
\prod_i \si^x_i \prod_i s_{i,i+1} =1 
\end{align}
\ie we include only even $Z_2$-charge states in our  many-body Hilbert space.
The model \eqn{CZXb}, together with the $Z_2$-even Hilbert space, describes the
boundary of the 2+1D DS topological order.  Such a 1+1D theory has a
non-invertible gravitational anomaly described by 2+1D DS
topological order.

Now we see that using the partition functions $Z_{a_x,a_t}$ of the model
\eqn{CZXb} with different $Z_2$-symmetry twists, we can construct the four
partition functions for the gapless boundary of 2+1D DS topological order.  For
example, the partition function of the model \eqn{CZXb} in the even $Z_2$
charge sector, $\frac{Z_{1,1}+Z_{1,-1}}{2}$, corresponds to the partition
function for the boundary of the DS topological order without an insertion,
$Z(\tau,\bar\tau,\one)$.  Note that the DS topological order has four types of
excitations: trivial excitation $\one$, semion $s$, conjugate semion $s^*$, and
topological boson $b$.  Thus the boundary has four partition functions
$Z(\tau,\bar\tau,\one)$, $Z(\tau,\bar\tau,s)$, $Z(\tau,\bar\tau,s^*)$, and
$Z(\tau,\bar\tau,b)$, which are given by
\begin{align}
\label{ZbDS}
 Z(\one) & = \frac{Z_{1,1}+Z_{1,-1}}{2}, \ \
 Z(s)  = \frac{Z_{-1,1}+Z_{-1,-1}}{2}, 
\nonumber \\
 Z(b)  &= \frac{Z_{1,1}-Z_{1,-1}}{2},\ \
 Z(s^*)  = \frac{Z_{-1,1}-Z_{-1,-1}}{2}.
\end{align}
The 2+1D DS topological order
is characterized by (in the basis of $\one,s,s^*, b$)
\begin{align}
\label{STDS}
  T^\text{top}_\text{DS}&=
\begin{pmatrix}
    1&0&0&0\\
    0&\ii&0&0\\
    0&0&-\ii&0\\
    0&0&0&1
  \end{pmatrix} ,
&
  S^\text{top}_\text{DS}&=\frac12 \begin{pmatrix}
    1&1&1&1\\
    1&-1&1&-1\\
    1&1&-1&-1\\
    1&-1&-1&1
  \end{pmatrix}.
\end{align}
Using the above $S^\text{top}_\text{DS}$ and $T^\text{top}_\text{DS}$ and the
modular transformations of $u1_{16}$ in Appendix \ref{U1CFT}, we can explicitly
check that the four boundary partition functions \eqn{ZbDS} indeed satisfy the
modular covariance \eqn{ZST}.

\section{Non-invertible gravitational anomaly and ``non-locality'' of
Hilbert space}

In \Ref{W1313}, it was stressed that the 't Hooft anomaly of a global symmetry
in a theory is not an obstruction for the theory to have an ultra-violate (UV)
completion (\ie to have a lattice realization).  Such an anomalous theory can
still be realized on a lattice, however, the global symmetry has to be realized
as an \emph{non-on-site} symmetry in the lattice model.

In this section, we would like to propose that a non-invertible gravitational
anomaly in a theory is not an obstruction for the theory to have a UV
completion.  The anomalous theory can still be realized on a lattice, if there
is no perturbative gravitational anomaly.  However, the Hilbert space of UV
theory $\cV$ is not given by the lattice Hilbert space $\cV_\text{latt}$: $\cV
\neq \cV_\text{latt}$.  

The lattice Hilbert space $\cV_\text{latt}$ has a tensor product decomposition 
\begin{align}
\cV_\text{latt} = \otimes_i \cV_i 
\end{align} 
where $\cV_i$ is the Hilbert space for each lattice site.  We call the  Hilbert
space with such a tensor product decomposition as \emph{local} Hilbert space.
A system with such a local  Hilbert space is free of gravitational anomaly, by
definition.

In contrast to an anomaly-free theory, the UV completion of a theory with
non-invertible gravitational anomaly does not have a local Hilbert space (\ie
with the above tensor product decomposition).  In other words, a non-invertible
gravitational anomaly is not an obstruction to have a UV completion, but for
the UV completion to have a local Hilbert space.  This understanding of
non-invertible gravitational anomaly is supported by the example discussed in
the last section.

In last section, we pointed out the boundary of 2+1D $Z_2$ topological order
(which has a non-invertible gravitational anomaly) has a UV completion
described by a lattice model \eq{tIsing}, with a constraint on the Hilbert
space $\prod_i \si^z_i =1$.  It is the constraint $\prod_i \si^z_i =1$ that
makes the Hilbert space non-local.

Let us describe the above result using a more physical reasoning.  One boundary
of 2+1D $Z_2$ topological order has a single type of topological excitations
$e$, which have a mod 2 conservation.  The Hilbert space always has an even
number of $e$ particles.  On the other hand, when there is no  $e$-particle
excitations, the boundary ground state is not degenerate.  Here we like to
point out that \frmbox{the even-particle constraint (\ie $Z_2$ fusion) plus
the non-degeneracy of the ground state is a sign of 1+1D non-invertible
gravitational anomaly.}

The example in the last section supports such a claim.  The even-particle
constraint is imposed by $\prod_i \si^z_i =1$.  The non-degenerate ground state
is given by $\otimes_i |\si^z=1\>$.  Such a theory describes a boundary of 2+1D
$Z_2$ topological order and has a non-invertible anomaly.  The Ising model may
also in the symmetry breaking phase.  Due to the constraint $\otimes_i
|\si^z=1\>$, the symmetry breaking phase also has a unique ground state
$\otimes_i |\si^x=1\> + \otimes_i |\si^x=-1\>$.  In such a symmetry breaking
phase, there are always an even number of domain walls, that correspond to an
even  number of topological excisions.

On the other hand, even-particle constraint plus two-fold degenerate ground
states will lead to an anomaly-free theory.  We can consider an Ising model in
symmetry breaking phase and without the constraint on Hilbert space.  Such a
phase has two-fold degenerate ground states and the number of domain walls
(which correspond to the $e$-particles) is always even. Thus even-particle
constraint plus two-fold degenerate ground states can be realized by a lattice
model with local Hilbert space and is thus anomaly-free.

There is also a mathematical way to understand the above claim.  The
$e$-particles with mod 2 conservation in 1+1D can be described by a fusion
category with a $Z_2$ fusion ring.  There are only two different fusion
categories with a $Z_2$ fusion ring, both have 1+1D non-invertible anomaly.
One fusion category describes the boundary of 2+1D $Z_2$ topological order, and
the other describes the boundary of DS topological order.
Both non-invertible anomalies can be described by the following Ising model 
\begin{align}
 H = -\sum_i \si^z_i \si^z_{i+1}
\end{align}
but with different constraints on Hilbert space.  The  anomaly corresponds to
the $Z_2$ topological order has a constraint $\prod_i \si^x_i =1$, and the
anomaly corresponds to the DS topological order has a constraint $\prod_i
\si^x_i \prod_i s_{i,i+1}=1$ (see \eqn{Xs}).

\section{Systematical search of gapped and gapless boundaries of a 2+1D
topological order}

\subsection{Boundaries of 2+1D topological order}

In this section, we want to systematically find gapped and gapless boundaries
of a 2+1D topological order by solving  \eqn{ZST} from the data $ S^\text{top},
T^\text{top}$ of the bulk topological order.  This is a generalization of
finding possible 1+1D critical theories via finding modular invariant partition
functions.   Note that, regardless of whether the boundary is gapped or gapless,
it always has the same  anomaly characterized by the bulk topological order.

To solve \eqn{ZST}, we may start with a CFT with partition
functions $Z_\text{bdy}(\tau,\bar\tau,I)$, which transform as
\begin{align}
 T_{IJ} Z_\text{bdy}(\tau,\bar\tau,J) &= Z_\text{bdy}(\tau+1,\bar\tau+1,I), 
\nonumber\\
 S_{IJ} Z_\text{bdy}(\tau,\bar\tau,J) &= Z_\text{bdy}(-1/\tau,-1/\bar\tau,I).
\end{align}
We then construct $Z(\tau,\bar\tau,i)$ via
\begin{align}
\label{ZMZ}
Z(\tau,\bar\tau,i) = M_{iI} Z_\text{bdy}(\tau,\bar\tau,I). 
\end{align}
Now \eqn{ZST} becomes
\begin{align}
M_{iI} Z_\text{bdy}(\tau+1,\bar\tau+1,I) &= M_{iI} T_{IJ}  Z_\text{bdy}(\tau,\bar\tau,J)
\\
&=  (T^\text{top})_{ij} M_{jJ} Z_\text{bdy}(\tau,\bar\tau, J)
\nonumber 
\end{align}
We see that $M_{iI}$ must satisfy
\begin{align}
\label{MST}
 M_{iI} &= (T^\text{top})_{ij} T^*_{IJ} M_{jJ}, & 
 M_{iI} &= (S^\text{top})_{ij} S^*_{IJ} M_{jJ}. 
\end{align}

We also note that, for a fixed $i$, $Z(\tau,\bar\tau,i)$ can be zero,
indicating the always presence of gapped excitations on the boundary. 
$Z(\tau,\bar\tau,i)$ can also be a $\tau$-independent positive integer.  It
means that the ground states are gapped and have a degeneracy given by
$Z(\tau,\bar\tau,i)$. Otherwise, $Z(\tau,\bar\tau,i)$ has an expansion
\begin{align}
\label{Zq}
&
Z(\tau,\bar\tau,i) = 
q^{-\frac{c}{24}} 
\bar q^{-\frac{\bar c}{24}} 
\sum_{n,\bar n=0}^\infty D_{n,\bar n}(i) 
q^{n+h_i}
\bar q^{\bar n+\bar h_i}
, 
\nonumber\\
&
q = \ee^{\ii 2\pi \tau},\ \ \ \
D_{n,\bar n}(i) = \text{non-negative integer}. 
\end{align}
where $(h_i,\bar h_i)$ are the scaling dimensions for the type-$i$ topological
excitation.  Such an expansion describes the many-body spectrum of the gapless
boundary of the disk $D^2_i$, with a type-$i$ topological excitation at the
center of the disk.  Here the subscript $i$ in $D^2_i$ indicates the type-$i$
excitation on the disk.  Let us assume the boundary $S^1=\prt D^2_i$ has a
length $L$.  Then $D_{n,\bar n}(i)$ is number of many-body states on $D^2_i$
with energy $(n+h_i+\bar n+\bar h_i)\frac{2\pi}{L}$, and momentum $(n+h_i-\bar
n-\bar h_i)\frac{2\pi}{L}$. Here we have assumed that velocity of the gapless
excitations is $v=1$.  Thus $D_{n,\bar n}(i)$ are non-negative integers. 

Also $D_{0,0}(i)$ is the ground state degeneracy on the boundary of the disk
$D^2_i$. Since the boundary can be gapless, the ground state degeneracy needs
to be defined carefully.  Here, we view two energy levels with an energy
difference of order $2\pi/L$ as non-degenerate.  We  view two energy levels
with an energy difference smaller than $(2\pi/L)^\al$, $\al >1$, as degenerate.
It is in this sense we define the ground state degeneracy $D_{0,0}(i)$ for
a gapless system in $L\to \infty$ limit.  We believe that the ground state
degeneracy on disk $D^2$ is always 1.  Therefore, we like to impose a
nondegeneracy condition on the boundary $D_{0,0}(\one)=1$.  
$Z_\text{bdy}(\tau,\bar\tau, I)$ satisfies a similar quantization condition.

From \eqn{ZMZ}, we see that $M_{iI}$ is the multiplicity of the number of
energy levels in the many-body spectrum of the boundary theory.  Therefore, for
a fixed $i$, if $M_{iI}\neq 0$, then
\begin{align}
\label{Mq}
& 
M_{iI} \text{ are quantized to make } 
D_{n,\bar n}(i) 
\nonumber\\
& 
\text{ to be non-negative
integer and } D_{0,\bar 0}(\one)=1.
\end{align}
In practice, to find $M_{iI}$, we may compute the eigenvectors of $
T^\text{top}\otimes T^* + S^\text{top}\otimes S^* $ with eigenvalue
2, that satisfy the above quantization condition.

\subsection{$Z_2$ topological order}

To find a CFT that describes a boundary of 2+1D $Z_2$ topological order, we
need to solve \eqn{ZST} with $S^\text{top},\ T^\text{top}$ given by
\eqn{Z2STmat} that characterize the 2+1D $Z_2$ topological order.  Let us first
try to find gapped boundaries by choosing $Z_\text{bdy}(\tau,\bar\tau)=1$, the partition
function of a trivial gapped 1+1D state.  Now \eqn{MST} reduces to
\begin{align}
 Z(i) =  (T^\text{top}_{Z_2})_{ij} Z(j), \ \ \ \ 
 Z(i) =  (S^\text{top}_{Z_2})_{ij} Z(j).
\end{align}
So we need to find common eigenvectors of $S^\text{top}_{Z_2}$ and
$T^\text{top}_{Z_2}$, both with eigenvalue 1.  We also require the solutions to
satisfy the quantization condition \eqn{Mq}, \ie the components of the
solutions are all non-negative integers. The condition $D_{0,0}(\one)=1$
becomes $Z(\one)=1$. This agrees with the fact that the ground state of 2+1D
$Z_2$ topological order on a disk $D^2$ is non-degenerate if there is no
accidental degeneracy.  This can be achieved by finding eigenvectors of
$S^\text{top}_{Z_2}+T^\text{top}_{Z_2}$ that satisfy \eqn{Mq}.

We find that $S^\text{top}_{Z_2}+T^\text{top}_{Z_2}$ has two eigenvectors with
eigenvalue 2, given by
\begin{align}
\label{Z2gapped}
 ( Z_m(i) ) = (1,0,1,0),\ \ \ \  ( Z_e(i) ) = (1,1,0,0) ,
\end{align}
where $i=(\one,e,m,f)$.  They are the only two non-negative integral
eigenvectors with $Z(\one)=1$.  Thus the 2+1D $Z_2$ topological order has only
two types of gapped boundaries, an $e$ condensed boundary described by $Z_e(i)$
and an $m$ condensed boundary described by $Z_m(i)$.\cite{LWW1414}

If we choose $Z_\text{bdy}(\tau,\bar\tau,I)$ to be the partition functions (the
characters) of $\text{Is}\otimes \overline{\text{Is}}$ CFT (see Appendix
\ref{minCFT}), then $S,T$ will be $9\times 9$ matrices:  
\begin{align}
 S_{\text{Is}\otimes \overline{\text{Is}}} = S^*_\text{Is}\otimes S_\text{Is}, \ \ \
 T_{\text{Is}\otimes \overline{\text{Is}}} = T^*_\text{Is}\otimes T_\text{Is}, \ \ \
\end{align}
where $S_\text{Is}, T_\text{Is}$ are given in \eqn{IsingST}.  We find
eigenvalue 2 for $ T^\text{top}_{Z_2}\otimes T_{\text{Is}\otimes
\overline{\text{Is}}}^* + S^\text{top}_{Z_2}\otimes S_{\text{Is}\otimes
\overline{\text{Is}}}^* $ to be 3-fold degenerate.  We obtain the following
three solutions of \eqn{ZST}
\begin{align}
\begin{pmatrix}
 Z(\tau,\bar\tau,\one)\\
 Z(\tau,\bar\tau,e)\\
 Z(\tau,\bar\tau,m)\\
 Z(\tau,\bar\tau,f)\\
\end{pmatrix} 
&=
\begin{pmatrix}
|\chi_\one(\tau)|^2+|\chi_\psi(\tau)|^2+|\chi_{\si}(\tau)|^2\\
|\chi_\one(\tau)|^2+|\chi_\psi(\tau)|^2+|\chi_{\si}(\tau)|^2\\
0\\
0\\
\end{pmatrix} ,
\\
\begin{pmatrix}
 Z(\tau,\bar\tau,\one)\\
 Z(\tau,\bar\tau,e)\\
 Z(\tau,\bar\tau,m)\\
 Z(\tau,\bar\tau,f)\\
\end{pmatrix} 
&=
\begin{pmatrix}
|\chi_\one(\tau)|^2+|\chi_\psi(\tau)|^2+|\chi_{\si}(\tau)|^2\\
0\\
|\chi_\one(\tau)|^2+|\chi_\psi(\tau)|^2+|\chi_{\si}(\tau)|^2\\
0\\
\end{pmatrix} ,
\\
\label{Z2b3}
\begin{pmatrix}
 Z(\tau,\bar\tau,\one)\\
 Z(\tau,\bar\tau,e)\\
 Z(\tau,\bar\tau,m)\\
 Z(\tau,\bar\tau,f)\\
\end{pmatrix} 
&=
\begin{pmatrix}
|\chi_\one(\tau)|^2+|\chi_\psi(\tau)|^2\\
|\chi_{\si}(\tau)|^2\\
|\chi_{\si}(\tau)|^2\\
\chi_\one(\tau) \bar \chi_\psi(\tau) +\chi_\psi(\tau) \bar \chi_\one(\tau) \\
\end{pmatrix} ,
\end{align}
that satisfy the quantization condition \eqn{Mq}.  

The first two solutions correspond to the two gapped boundaries of the 2+1D
$Z_2$ topological order induced by $e$ and $m$ condensation respectively, and
then stacking with a transverse Ising model at critical point.  So the first
two solutions are regarded as gapped boundaries.  Here we would like introduce
the notion of reducible boundary.  If the partition functions $Z(\tau,\bar\tau,i)$ of a
boundary has a form
\begin{align}
 Z(\tau,\bar\tau,i) = Z_\text{inv}(\tau,\bar\tau) Z'(\tau,\bar\tau,i),
\end{align}
then we say the boundary is reducible. We will call the boundary described by
$Z'(\tau,\bar\tau,i)$ as the reduced boundary.  Here $Z(\tau,\bar\tau,i)$ and $Z'(\tau,\bar\tau,i)$ are
partition functions satisfying \eq{ZST} and \eqn{Zq}, and $Z_\text{inv}(\tau,\bar\tau)$
is a modular invariant partition function satisfying \eqn{Zq}.  Noticing that
$|\chi_\one(\tau)|^2+|\chi_\psi(\tau)|^2+|\chi_{\si}(\tau)|^2$ is modular
invariant, so the first two boundaries are reducible and their reduced boundary
are gapped boundaries described by \eqn{Z2gapped}.

The third solution \eq{Z2b3} corresponds to an irreducible gapless boundary.
Now we like to consider the stability of such $c=\bar c=\fh$ gapless boundary.
But before that, we want review the stability of the critical point of
transverse Ising model described by
\begin{align}
Z_\text{Is}(\tau,\bar\tau)=|\chi_\one(\tau)|^2+|\chi_\psi(\tau)|^2+|\chi_{\si}(\tau)|^2 .
\end{align}
From the above partition function, we see that there are two relavant
operators: $\bar \psi \psi$ with scaling dimension $(h,\bar h)=(\frac12,
\frac12)$, and $\bar \si \si$ with scaling dimension $(h,\bar h)=(\frac1{16},
\frac1{16})$.  Among the two, $\bar \si \si$ is odd under the $Z_2$ symmetry of
the transverse Ising model.

Similarly, to examine the  stability of the gapless boundary \eq{Z2b3}, we
examine the partition function $Z(\tau,\bar\tau,\one)$.  We do not consider
other partition functions, since the partition function $Z(\tau,\bar\tau,\one)$
describes the physical boundary of Fig. \ref{bdyblk} without the insertion of
the world-line.  From $Z(\tau,\bar\tau,\one)$, we see that gapless boundary
\eq{Z2b3} has only one relavant operator $\bar \psi \psi$ with scaling
dimension $(h,\bar h)=(\frac12, \frac12)$.  So the gapless boundary \eq{Z2b3}
can be the phase transition point between two gapped boundaries.  In fact,
according to the discussion in Section \ref{gapless}, the third gapless
boundary is the critical transition point between the gapped $e$ condensed
boundary and the $m$ condensed boundary.

We can also use the characters $\chi^{m4}_h$ of the $(4,5)$ minimal model (or tricritical Ising model \cite{francesco2012conformal}) with central charge $c=\bar c=\frac{7}{10}$, to
construct the boundary partition functions $Z_\text{bdy}(\tau,\bar\tau,I)$ that
have the $Z_2$ non-invertible anomaly (\ie satisfy \eqn{ZST}).  We obtain
\begin{align}
\begin{pmatrix}
Z(\one) \\ Z( e) \\ Z( m) \\ Z( f) \\
\end{pmatrix}=
\begin{pmatrix}
|\chi^{m4}_0|^2 +  |\chi^{m4}_\frac{1}{10}|^2 +  |\chi^{m4}_\frac{3}{5}|^2 +  |\chi^{m4}_\frac{3}{2}|^2 \\
|\chi^{m4}_\frac{7}{16}|^2 +  |\chi^{m4}_\frac{3}{80}|^2 \\
|\chi^{m4}_\frac{7}{16}|^2 +  |\chi^{m4}_\frac{3}{80}|^2 \\
\chi^{m4}_0 \bar\chi^{m4}_\frac{3}{2} +  \chi^{m4}_\frac{1}{10} \bar\chi^{m4}_\frac{3}{5} +  \chi^{m4}_\frac{3}{5} \bar\chi^{m4}_\frac{1}{10} +  \chi^{m4}_\frac{3}{2} \bar\chi^{m4}_0 \\
\end{pmatrix}
\end{align}
If we choose $Z_{\text{bdy}}(\tau,\bar \tau,I)$ to be built from the characters
of $u(1)_M\otimes \bar{u(1)}_M$ CFT, we obtain the following simple solution of
gapless boundary
\begin{align}
\begin{pmatrix}
Z(\tau,\bar \tau; \one) \\ Z(\tau,\bar \tau; e) \\ Z(\tau,\bar \tau; m) \\ Z(\tau,\bar \tau; f) \\
\end{pmatrix}=
\begin{pmatrix}
\left|\chi_0^{u1_4}\right|^2+\left|\chi_2^{u1_4}\right|^2\\
\left|\chi_1^{u1_4}\right|^2+\left|\chi_3^{u1_4}\right|^2\\
\chi_1^{u1_4}\bar \chi_3^{u1_4}+\chi_3^{u1_4}\bar \chi_1^{u1_4}\\
\chi_0^{u1_4}\bar \chi_2^{u1_4}+\chi_2^{u1_4}\bar \chi_0^{u1_4}\\
\end{pmatrix}
\end{align}
Note that $Z(\tau, \bar \tau, e)$ and $Z(\tau,\bar \tau, m)$ are no longer
identical, but differ by a charge conjugation, whose action induce on the
characters is $C: \chi_i^{u1_M}\bar \chi_j^{u1_M}\rightarrow \chi_i^{u1_M}\bar
\chi_{M-j}^{u1_M}$.

\subsection{Double-semion topological order}

To find gapped boundaries of  2+1D DS topological order, we need
to solve
\begin{align}
 Z(i) =  (T^\text{top}_\text{DS})_{ij} Z(j), \ \ \ \ 
 Z(i) =  (S^\text{top}_\text{DS})_{ij} Z(j),
\end{align}
where $T^\text{top}_\text{DS},S^\text{top}_\text{DS}$ are given by \eqn{STDS}.
We find that $S^\text{top}_\text{DS}+T^\text{top}_\text{DS}$ has only one
eigenvector with
eigenvalue 2, given by
\begin{align}
\label{DSgapped}
 ( Z_b(i) ) = (1,0,0,1),
\end{align}
where $i=(\one,s,s^*,b)$.  Thus the 2+1D DS topological order has
only one type of gapped boundary, a $b$ condensed boundary.\cite{LWW1414}

Next, we consider possible gapless boundaries of DS topological order
described by $\text{Is}\otimes \overline{\text{Is}}$ CFT, by solving \eqn{MST}
for solutions satisfying \eqn{Mq}.  We find only one eigenvector for $
T^\text{top}_\text{DS}\otimes T_{\text{Is}\otimes \overline{\text{Is}}}^* +
S^\text{top}_\text{DS}\otimes S_{\text{Is}\otimes \overline{\text{Is}}}^* $
with eigenvalue $2$.
We obtain the following unique
solution of \eqn{ZST}
\begin{align}
\begin{pmatrix}
 Z(\tau,\bar\tau,\one)\\
 Z(\tau,\bar\tau,s)\\
 Z(\tau,\bar\tau,s^*)\\
 Z(\tau,\bar\tau,b)\\
\end{pmatrix} 
&=
\begin{pmatrix}
|\chi_\one(\tau)|^2+|\chi_\psi(\tau)|^2+|\chi_{\si}(\tau)|^2\\
0\\
0\\
|\chi_\one(\tau)|^2+|\chi_\psi(\tau)|^2+|\chi_{\si}(\tau)|^2\\
\end{pmatrix} 
\end{align}
Such a solution corresponds to the gapped boundary of 2+1D DS
topological order, and then stacking with a transverse Ising model at critical
point.  So this solution is regarded as a gapped boundary.  There is no
irreducible gapless boundary described by $\text{Is}\otimes
\overline{\text{Is}}$.  

Actually, we can obtain an even stronger result \frmbox{2+1D DS topological
order has no irreducible gapless boundary with central charge $c=\bar
c<\frac{25}{28}$.}   This result is obtained by realizing that the DS anomalous
partition functions, for irreducible gapless boundary, has a non-zero component
$Z(\tau,\bar\tau,s)$.  Otherwise, the gapless boundary can be viewed as a
gapped boundary stacked with an anomaly-free 1+1D CFT.  The condition \eq{ZST}
for the $T^\text{top}$-transformation requires that the excitations in the
partition function has topological spin $h-\bar h=\frac14$ mod 1.  This
constraint the central charge of the anomalous CFT.  If the CFT has a central
charge $c=\bar c<1$, then the boundary CFT must be given by a
chiral-anti-chiral minimal model $C^\text{ft}_{p,p+1}\times \bar
C^\text{ft}_{p,p+1}$.  The topological spin for the operators in such CFT is
given by $s_{r,s,r',s'}=h_{r,s}-h_{r',s'}$ (see \eqn{chrs1}).  We find that,
for $p<7$, $s_{r,s,r',s'}$ cannot be $\frac14$ mod 1.  Thus the condition
\eqn{ZST} cannot by satisfied for $T^\text{top}$ transformation.

Last, we consider possible gapless boundary theories of DS topological order described
by $u1_M\otimes \overline{u1}_M$ CFT, by solving \eqn{MST} for solutions
satisfying \eqn{Mq}.  This includes many cases, one for each choice of $M$. So
we need to consider each case separately.

For $M=16$, we have found an irreducible gapless boundary described by
$u1_{16}\otimes \overline{u1}_{16}$ CFT:
\begin{align}
\label{DSu1-16}
\begin{pmatrix}
 Z(\one)\\
 Z(s)\\
 Z(s^*)\\
 Z(b)\\
\end{pmatrix} 
&=
\begin{pmatrix}
\sum_{i=0}^3 |\chi_{4i}^{u1_{16}}|^2
+\sum_{i=0}^3 \chi_{4i+10}^{u1_{16}} \bar\chi_{4i+2}^{u1_{16}} 
\\
\sum_{i=0}^3 \chi_{4i+1}^{u1_{16}} \bar\chi_{4i+5}^{u1_{16}} 
+\sum_{i=0}^3 \chi_{4i+3}^{u1_{16}} \bar\chi_{4i+15}^{u1_{16}} 
\\
\sum_{i=0}^3 \chi_{4i+3}^{u1_{16}} \bar\chi_{4i+7}^{u1_{16}} 
+\sum_{i=0}^3 \chi_{4i+1}^{u1_{16}} \bar\chi_{4i+13}^{u1_{16}} 
\\
\sum_{i=0}^3 |\chi_{4i+2}^{u1_{16}}|^2
+\sum_{i=0}^3 \chi_{4i+8}^{u1_{16}} \bar\chi_{4i}^{u1_{16}} 
\\
\end{pmatrix} 
\end{align}
However, it is not clear if there are other irreducible gapless boundaries
described by $u1_{16}\otimes \overline{u1}_{16}$ CFT.

From the partition function $Z(\tau,\bar\tau,\one)$, we see that there is an
relevant operator with scaling dimension $(h,\bar h) = ( \frac{4^2}{2\times
16}, \frac{4^2}{2\times 16}) = ( \frac{1}{2}, \frac{1}{2}) $.  So the gapless
boundary \eq{DSu1-16} is unstable.  It describes the transition point between
two gapped phases in \eqn{CZXb}.  One gapped phase for $U>0$ and other gapped
phase for $U<0$.  The gapless critical point is described by $U=0$.

For $M=4$, we find that there is no irreducible gapless boundary described by
$u1_4\otimes \overline{u1}_4$ CFT.

For $M=2$, we find that there is only one irreducible gapless boundary described
by $u1_2\otimes \overline{u1}_2$ CFT:
\begin{align}
\label{DSu1-2}
\begin{pmatrix}
 Z(\tau,\bar\tau,\one)\\
 Z(\tau,\bar\tau,s)\\
 Z(\tau,\bar\tau,s^*)\\
 Z(\tau,\bar\tau,b)\\
\end{pmatrix} 
&=
\begin{pmatrix}
|\chi_0^{u1_2}|^2 \\
\chi_1^{u1_2} \bar\chi_0^{u1_2} \\
\chi_0^{u1_2} \bar\chi_1^{u1_2} \\
|\chi_1^{u1_2}|^2 \\
\end{pmatrix} 
\end{align}

There is no other irreducible gapless boundary described by $u1_2\otimes
\overline{u1}_2$ CFT. But there is a reducible
gapless boundary described by
\begin{align}
\begin{pmatrix}
 Z(\tau,\bar\tau,\one)\\
 Z(\tau,\bar\tau,s)\\
 Z(\tau,\bar\tau,s^*)\\
 Z(\tau,\bar\tau,b)\\
\end{pmatrix} 
&=
\begin{pmatrix}
|\chi_0^{u1_2}|^2 + |\chi_1^{u1_2}|^2 \\
0 \\
0 \\
|\chi_0^{u1_2}|^2 + |\chi_1^{u1_2}|^2 \\
\end{pmatrix} 
\end{align}
which is a stacking of a gapped boundary described by \eqn{DSgapped}
and the CFT for spin-1/2 Heisenberg chain.

From the partition function $Z(\tau,\bar\tau,\one)$, we find that the
irreducible boundary \eq{DSu1-2} has no relevant operator.  It has only several
marginal operators, such as $\bar J J$, $ \ee^{\pm \ii \sqrt 2\phi} \ee^{\pm
\ii \sqrt 2\bar \phi}$ with scaling dimension $(h,\bar h)=(1,1)$.  Here $J$ is
the $U(1)$ current operator and $\ee^{\pm \ii \sqrt 2\phi}$ are $U(1)$-charged
operators. Those operators can be marginally relavent.  If there is only one
marginally relavent operator $g \hat O$ in the Hamiltonian, the renomalization
group (RG) flow of the coupling constant $g$ is given by
\begin{align}
 \frac{\dd g}{\dd \bt} = \al g^2.
\end{align} 
We see that regardless the sign of $\al$, there is a finite region of $g$
where $g$ flows to zero.  In this case, the CFT can be stable.  When there are
many marginally relevant operators  $g_i \hat O_i$, RG flow of the coupling
constants $g_i$ is given by\cite{cardy1996scaling}
\begin{align}
 \frac{\dd g_i}{\dd \bt} = \al_{ijk} g_j g_k .
\end{align} 
In Appendix \ref{RG}, we discuss the above RG equation in more details and show
that generic coupling constants $g_i$ always flow to infinite.  Thus, the CFT
is unstable, and the 2+1D DS topological order always has a gapped boundary
without fine tuning.

We like remark that from this gapless boundary of DS topological order, and
apply the relations \eq{ZbDS}, we can find another galpess boundary theory
of $\ZZ_2$ SPT, whose partition function is given by
\begin{align}
\label{Z2SPT}
Z_\text{$Z_2$-SPT}(q,\bar q)=\sum_{i=0}^1 |\chi_i^{u1_2}(q)|^2 
\end{align}
which is different from \eqn{ZXY}.
%without symmetry twist.  with primary fields of dimension $( h_i,\bar
%h_i)=\left(\frac{i^2}{4},\frac{i^2}{4}\right)$. 
The partition function \eq{Z2SPT} can also be rewritten as
\begin{align}
Z_\text{$Z_2$-SPT}(q,\bar q)=|\eta (q)|^{-1}\sum_{i=0}^1 q^{\frac{1}{2}a^2}\bar q^{\frac{1}{2}b^2}
\end{align}
where $(a,b)$ form a lattice $\Gamma^{u1_2}$,
\begin{align}
(a,b)=\frac{1}{\sqrt{2}}(l+m,l-m),\quad l,m\in\ZZ
\end{align}
The $\ZZ_2$ charges of the vectex operators in $|\chi_i^{u1_2}|^2$ is $i\mod 2$, or $(l+m)\mod 2$ on the lattice.

From the $Z_2$-even partition function $|\chi_0^{u1_2}(q)|^2$, we find that the
gapless boundary \eq{Z2SPT} has no $Z_2$-even relevant operator.  However, as
mentioned above, there may be many marginally relavent  operators, and it is
not clear if the gapless boundary \eq{Z2SPT} of 2+1D $Z_2$-SPT order is
perturbatively stable or not.  In some previous studies, a gapless boundary
\eq{ZXY} for the same 2+1D $Z_2$-SPT order is found to be perturbatively
unstable  against $Z_2$ symmetric perturbations,\cite{CW1235,LG1209} via
relavent perturbations (with total scaling dimension less than 2).  In this
paper, we found a gapless boundary of $Z_2$-SPT state \eq{Z2SPT}, which is more
stable against $Z_2$ symmetric perturbations, in the sense that the instability
only come from potentially marginally relevant operators (with total scaling
dimension equal to 2).

\subsection{Single-semion topological order}

There is a close relative of 2+1D DS topological order -- 2+1D single-semion
(SS) topological order, which has only two types of excitations: trivial
excitation $\one$ and semion $s$.  The 2+1D SS topological order can be
realized by $\nu=1/2$ bosonic Laughlin state.

Let us describe the data that characterizes the 2+1D SS topological order.  The
topological spins and the quantum dimensions of $\one$ and $s$ are $(s_\one,
s_s)=(0,\frac14)$ and $(d_\one, d_s)=(1,1)$.  The topological $
S^\text{top}_\text{SS}, T^\text{top}_\text{SS}$ matrices are
\begin{align}
 T^\text{top}_\text{SS} &= \ee^{-\ii  \frac{2\pi}{24}}
\begin{pmatrix}
 1 & 0\\
 0 & \ee^{\ii  \frac{2\pi}4}
\end{pmatrix}
\nonumber\\
 S^\text{top}_\text{SS} &= \frac{1}{\sqrt 2}
\begin{pmatrix}
 1 & 1\\
 1 &-1\\
\end{pmatrix}
\end{align}

To obtain the possible boundaries of 2+1D SS topological order, we just need
to solve \eqn{ZST}. We find a simple boundary described by the following
partition function (in terms of $u1_2$ characters \eq{u1chi})
\begin{align}
\begin{pmatrix}
 Z(\tau,\bar\tau,\one)\\
 Z(\tau,\bar\tau,s)\\
\end{pmatrix} 
&=
\begin{pmatrix}
\chi_0^{u1_2}(\tau) \\
\chi_1^{u1_2}(\tau) \\
\end{pmatrix} 
\end{align}
The 1+1D theory described by the above partition functions has both
perturbative and global gravitational anomaly.

\subsection{Fibonacci topological order}

Another simple 2+1D topological order is the Fibonacci topological order.  It
is characterized by the following topological data. The central charge is $\frac{14}{5} \mod 8$. There are two types of
excitations $\one$ and $\ga$.  Their topological spins and the quantum
dimensions are $(s_\one, s_\ga)=(0,\frac 25)$ and $(d_\one, d_s)=(1,\phi)$,
where $\phi=\frac{\sqrt 5+1}{2}$, the golden ratio.  The topological $ S^\text{top}_\text{Fib},
T^\text{top}_\text{Fib}$ matrices are
\begin{align}
 T^\text{top}_\text{Fib} &= \ee^{-\ii  \frac{2\pi}{24}\frac{14}{5}}
\begin{pmatrix}
 1 & 0\\
 0 & \ee^{\ii  2\pi \frac 25}
\end{pmatrix}
\nonumber\\
 S^\text{top}_\text{Fib} &= \frac{1}{\sqrt {\phi+2}}
\begin{pmatrix}
 1 & \phi\\
 \phi &-1\\
\end{pmatrix}
\end{align}

Solving \eqn{ZST}, we can find several gapless boundary of the Fibonacci topological
order:
\begin{itemize}
\item $(G_2)_1$ CFT with central charge $(c, \bar c)= \left(\frac{14}{5}, 0\right)$, with the partition functions
\begin{align}
\label{FibG21}
&\begin{pmatrix}
 Z(\tau,\one)\\
 Z(\tau,\gamma)\\
\end{pmatrix} 
=
\begin{pmatrix}
\chi_0^{G2_1}(\tau) \\
\chi_1^{G2_1}(\tau) \\
\end{pmatrix} \nn\\
&=\ee^{-\ii \frac{2\pi}{24}\frac{14}{5}}\begin{pmatrix}
1+14q+42 q^2+O(q^3) \\
q^{\frac{2}{5}}\left(7+34 q+ 119q^2+O(q^3)\right)
\end{pmatrix}
\end{align}
where $\chi_i^{G2_1}(\tau)$ are the characters of level-1 $G_2$ current
algebra, see Appendix \ref{g2chi}.  The first multiplicity equaling $7$
appearing in $Z(\tau, \gamma)$ implies that when there is a Fibonacci anyon in
the bulk, the boundary has $7$-fold degeneracy. The degeneracy cannot be split
unless the anyon is moved to the boundary.

\item $su(2)_3\times u(1)_M$ CFT has a central charge
$c=\frac{9}{5}+1=\frac{14}{5}$ and $\bar c=0$.  The $su2_3$ CFT has four chiral
characters $\chi_j^{su2_3}$, labeled by the spin
$j=0,\frac{1}{2},1,\frac{3}{2}$ (see Appendix \ref{SU2CFT}) with $S,T$-matrices
\begin{align}
T_{su2_3}=&\ee^{-\ii \frac{2\pi}{24}\frac{9}{5}}\begin{pmatrix}
1 & 0 & 0 & 0 \\
0 & \ee^{\ii 2\pi\frac{3}{20}} & 0 & 0 \\
0 & 0 & \ee^{\ii 2\pi\frac{2}{5}} & 0 \\
0 & 0 & 0 & \ee^{\ii 2\pi\frac{3}{4}}
\end{pmatrix}\nn\\
S_{su2_3}=&\frac{1}{\sqrt{2 (\phi+2)}}\begin{pmatrix}
1 & \phi & \phi & 1 \\ \phi & 1 & -1 & -\phi \\ 
\phi & -1 & -1 & \phi \\ 1 & -\phi & \phi & -1 \\
\end{pmatrix}.
\end{align}
When $M=2$, we find a solution of \eqn{ZST}:
\begin{align}
\begin{pmatrix}
 Z(\tau,\one)\\
 Z(\tau,\gamma)\\
\end{pmatrix} 
&=
\begin{pmatrix}
\chi_0^{u1_2} \chi_0^{su2_3} + \chi_1^{u1_2} \chi_{\frac 32}^{su2_3} \\
\chi_1^{u1_2} \chi_{\frac 12}^{su2_3} + \chi_0^{u1_2} \chi_1^{su2_3} \\
\end{pmatrix}. \label{Fibsu23u12}
\end{align}
In fact, we find the expansion of the $Z(\tau,i)$ in \eqn{Fibsu23u12} in terms
of modular parameter $q=\ee^{\ii 2\pi \tau}$ to be the same as that of
\eqn{FibG21}. 
\item The same result also arises in $su(2)_{28}$ with $c=\frac{14}{5}$,
\begin{align}
Z(\tau,\one)=&\chi^{su2_{28}}_0+\chi^{su2_{28}}_{5}+\chi^{su2_{28}}_{9}+\chi^{su2_{28}}_{14}\nn\\
Z(\tau,\gamma)=&\chi^{su2_{28}}_3+\chi^{su2_{28}}_{6}+\chi^{su2_{28}}_{8}+\chi^{su2_{28}}_{11}.
\end{align}
and see Appendix \ref{SU2CFT} for explicit forms of characters.

\item $(E_8)_1\times \overline{(F_4)_1}$ CFT, with central charge $(c, \bar c)=\left(8, \frac{26}{5}\right)$, and $c-\bar c=\frac{14}{5}$. $(F_4)_{1}$ CFT has the $S,T$ matrices 
\begin{align}
T_{(F_4)_{1}}=\ee^{-\ii \frac{2\pi}{24} \frac{26}{5}}\begin{pmatrix}
1 & 0 \\ 0 & \ee^{\ii 2\pi \frac{3}{5}}
\end{pmatrix},\quad S_{(F_4)_{1}}=S_{\text{Fib}}^{\text{top}}.
\end{align}

Therefore, $(E_8)_1\times \overline{(F_4)_1}$ is also a gapless boundary of the
Fibonacci topological order.
\begin{align}
\begin{pmatrix}
 Z(\tau,\bar\tau,\one)\\
 Z(\tau,\bar\tau,\gamma)\\
\end{pmatrix} 
&=
\begin{pmatrix}
\chi^{(E_8)_1}(\tau)\bar \chi_0^{(F_4)_1}(\bar \tau) \\
\chi^{(E_8)_1}(\tau)\bar \chi_1^{(F_4)_1}(\bar \tau) \\
\end{pmatrix} 
\end{align}

%\item $\frac{(F_4)_1}{su2_8}\sim (G_2)_1$, with $c=\frac{26}{5}-\frac{12}{5}=\frac{14}{5}$.
%\item $su2_{28}\subset (G_2)_1$ with $c=\frac{14}{5}$

\end{itemize}

\section{Detect anomalies from 1+1D partition functions}

So far, we have discussed how to use anomaly to constraint the structure of
1+1D partition function.  In this section, we are going to consider a different
problem: given a partition function, how to determine its anomaly?  We have
mentioned that the 1+1D perturbative gravitational anomaly can be partially
detected via $q\to 0$ limit of partition function (see \eqn{pga}).  So here we
will concentrate on global gravitational anomalies.

Let us consider partition functions constructed using the characters of Ising
CFT:
\begin{align}
 Z_M(\tau,\bar\tau) = \sum_{i,j = 1,\psi, \si} \bar \chi_{i}(\bar\tau) M_{ij}  \chi_{j}(\tau)
\end{align}
Under modular transformation
$Z_M$ transforms as
\begin{align}
\label{Mtrans}
 Z_{M}(\tau+1,\bar\tau+1) &= Z_{M_T}(\tau,\bar\tau), &  M_T &= T_\text{Is}^\dag M T_\text{Is};
\nonumber\\
 Z_{M}(-1/\tau,-1/\bar\tau) &= Z_{M_S}(\tau,\bar\tau), &  M_S &= S_\text{Is}^\dag M S_\text{Is};
\end{align}
where $S_\text{Is},T_\text{Is}$ are given by \eqn{IsingST}.  

\begin{figure}[tb]
\begin{center}
\includegraphics[scale=0.6]{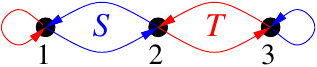} \end{center}
%Fig. 1
\caption{ 
The modular transformations on the partition functions $Z_{M_n}$, $n=\one,2,3$,
for a gapless boundary of a 2+1D $Z_2$ topological order.  For example, the two
red lines to the right represent the following $T$-transformations: $M_2\to
M_3: Z_{M_2}(\tau+1,\bar\tau+1)=Z_{M_3}(\tau,\bar\tau)$ and $M_3\to M_2:
Z_{M_3}(\tau+1,\bar\tau+1)=Z_{M_2}(\tau,\bar\tau)$.  The blue lines represent the
$S$-transformations.  The pattern of the transformations characterizes an 1+1D
non-invertible gravitational anomaly described by  2+1D $Z_2$ topological
order.
}
\label{Z2ST}
\end{figure}

Let us consider a particular partition function
\begin{align}
Z(\tau,\bar\tau,\one) &=Z_{M_\one}(\tau,\bar\tau), 
\nonumber\\
 M_\one &= \begin{pmatrix}
 1 & 0 & 0 \\
 0 & 1 & 0 \\
 0 & 0 & 0 \\
\end{pmatrix},
\end{align}
which is not modular invariant.  Starting from $M_\one$,
the modular transformations \eq{Mtrans} generate two other partition functions
described by
\begin{align}
 M_2 = \begin{pmatrix}
 \frac12 & \frac12 & 0 \\
 \frac12 & \frac12 & 0 \\
 0       & 0       & 1 \\
\end{pmatrix}
,\ \ \ \ \ \ \
 M_3 = \begin{pmatrix}
 \frac12 &-\frac12 & 0 \\
-\frac12 & \frac12 & 0 \\
 0       & 0       & 1 \\
\end{pmatrix}
\end{align}
The actions of modular transformations on $Z_{M_\one}$, $Z_{M_2}$, and
$Z_{M_3}$ are described by Fig. \ref{Z2ST}.  Such orbits of modular
transformations can be used to characterize the anomaly in the partition
function.  However, it is not clear if such a characterization is complete or
not, \ie  it is not clear if different anomalies always have different orbits.
However, the orbits in Fig. \ref{Z2ST} are consistent with the
1+1D anomaly described by 2+1D $Z_2$ topological order.  This is because the $
S^\text{top}_{Z_2}, T^\text{top}_{Z_2}$ transformations of the 2+1D $Z_2$
topological order \eq{Z2STmat}, when acting on
\begin{align}
 |\one\> \equiv 
\begin{pmatrix}
 1\\
0\\
0\\
0\\
\end{pmatrix}
\end{align}
will generate
\begin{align}
|2\> &\equiv \frac12
\begin{pmatrix}
 1\\
1\\
1\\
1\\
\end{pmatrix} 
=\frac12 ( |\one\> + |e\> + |m\> + |f\> )
\nonumber\\
|3\> &\equiv \frac12
\begin{pmatrix}
 1\\
1\\
1\\
-1\\
\end{pmatrix} 
=\frac12 ( |\one\> + |e\> + |m\> - |f\> )
\end{align}
The actions of $ S^\text{top}_{Z_2}, T^\text{top}_{Z_2}$ on $|\one\>$, $|2\>$,
and $|3\>$ will generate the same orbits as in Fig. \ref{Z2ST}.

\begin{figure}[tb]
\begin{center}
\includegraphics[scale=0.6]{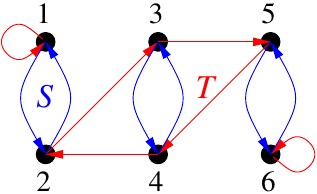} \end{center}
%Fig. 1
\caption{ 
The modular transformations on the partition functions $Z_n$,
$n=1,2,\cdots,6$, for a gapless boundary of 2+1D DS topological
order.  For example, the red lines in the middle represent the following
$T$-transformations: $Z_4\to Z_2: Z_4(\tau+1,\bar\tau+1)=Z_2(\tau,\bar\tau)$ and 
$Z_5\to Z_4: Z_5(\tau+1,\bar\tau+1)=Z_4(\tau,\bar\tau)$.  The blue lines represent the
$S$-transformations.  The pattern of the transformations characterizes an 1+1D
non-invertible gravitational anomaly described by  2+1D DS
topological order.
}
\label{DSST}
\end{figure}

We may also consider a partition function constructed using $u1_{16}$
characters:
\begin{align}
 Z(\tau,\bar\tau,\one)=
\sum_{i=0}^3 |\chi_{4i}^{u1_{16}}|^2
+\sum_{i=0}^3 \chi_{4i+10}^{u1_{16}} \bar\chi_{4i+2}^{u1_{16}}
\end{align}
Starting from $Z(\tau,\bar\tau,\one)=Z_1(\tau,\bar\tau)$, using modular transformations $S,T$ in
\eqn{u1ST}, we
can generate five additional partition functions $Z_n(\tau,\bar\tau)$, $n=2,3,4,5,6$.
Under the modular transformations $S,T$,  the  partition functions $Z_n(\tau,\bar\tau),\
n=1,\cdots,6$ change into each other. The orbits are described by Fig.
\ref{DSST}.  Such orbits are consistent with the 1+1D anomaly described by 2+1D
DS topological order

\section{Summary}

In this paper, we study non-invertible gravitational anomalies that correspond
to non-invertible topological orders in one higher dimension.  A theory with a
non-invertible anomaly can have many partition functions, which are linear
combinations of $N$ partition functions.  For 1+1D non-invertible anomaly, $N$
is the number types of the topological excitations in the corresponding 2+1D
topological order.  The anomalous 1+1D partition functions
$Z(\tau,\bar\tau,i)$, $i=1,\cdots,N$, are not invariant under the modular
transformation, but transform in a non-trivial way described by the modular
matrices $S^\text{top}_{ij}$ and $T^\text{top}_{ij}$ that characterize the
corresponding 2+1D topological order.  Similarly, anomalous theory on an
arbitrary close space-time manifold $M^d$ also has many partition functions
$Z(M^d,i)$, which transforms according to a representation $R_{M_d}$ of the
mapping class group $G_{M^d}$ of $M^d$.  The $G_{M^d}$ representation $R_{M_d}$
describes how the ground states of the corresponding $(d+1)$D topological order
transform on a spatial manifold $M^d$.  As an application of our theory of
non-invertible anomaly, we show that for 2+1D DS topological order, its
irreducible gapless boundary must have central charge $c=\bar c\geq
\frac{25}{28}$.

At the beginning of the paper, we mentioned that 't Hooft anomaly is an
obstruction to gauge a global symmetry.  However, if we include theories with
non-invertible anomaly, then even global symmetry with 't Hooft anomaly can be
gauged, which will result in a theory with a non-invertible anomaly.  This is
because a theory with 't Hooft anomaly can be realized as a boundary of SPT
state in one dimension higher, where the global symmetry is realized as an
on-site-symmetry. We can gauge the global on-site-symmetry in bulk and turn the
SPT state into a topologically ordered state.  The boundary of the resulting
topologically ordered state is the theory obtained by gauging the anomalous
global symmetry.  This connection between 't Hooft anomaly and non-invertible
gravitational anomaly allows us to use the theory on non-invertible
gravitational anomaly developed in this paper to systematically understand 't
Hooft anomaly and its effect on low energy properties.  Those issues will be
studied in \Ref{STanomaly}.

We like to thank Tian Lan and Samuel Monnier for many helpful discussions.
This research is partially supported by NSF grant DMS-1664412.

\appendix

\section{Characters of chiral CFTs}

\subsection{The minimal model CFT} \label{minCFT} \label{mm}

The chiral CFTs with central charge $c<1$ are called the minimal models.  They
are labeled by two integers $p,p'$ with $p,p'>2$ and an equivalence $(p,p')\sim
(p',p)$.  We demote those CFTs as $C^\text{ft}_{p,p'}$.  The central charge and
the dimensions of primary fields are given by
\begin{align}
 c &= 1- \frac{6(p-p')^2}{pp'}.
\nonumber\\
h_{r,s} &= \frac{(rp'-sp)^2-(p-p')^2}{4pp'},
\nonumber\\
 & 1\leq r \leq p-1, \ 1\leq s \leq p'-1,
\end{align}
which satisfy
\begin{align}
 h_{r,s}
=h_{p-r,p'-s}
=h_{p+r,p'+s}
\end{align}
The CFTs are unitary if and only if  $|p'-p|=1$.  In this case, the character
for the primary field  $(r,s)$ is given by
\begin{align}
 \chi_{r,s}(q) &= \frac{q^{h_{r,s}}}{\eta(q)}\sum_{n\in \Z} 
q^{n[(np+r)(p+1)-ps]} 
(1-q^{(2np+r)s}), 
\nonumber\\
\eta(q) &= q^{\frac 1{24}} \prod_{n=1}^\infty (1-q^n),
\ \ \ \ q=\ee^{2\ii \pi \tau},
\end{align}
where $\chi_{r,s}(q)= \chi_{p-r,p'-s}(q)$.  The $S$-matrix is
\begin{align}
S_{rs;\rho\sigma}=\sqrt{\frac{8}{pp'}}(-1)^{(1+s\rho+r\sigma)}
\sin\left(\pi \frac{p'}{p}r\rho\right)
\sin\left(\pi \frac{p}{p'}s\sigma\right)
\end{align}
For unitary minimal models $(p,p')=(p,p+1)$, we have
\begin{align}
\label{chrs1}
 c &= 1- \frac{6}{p(p+1)}.
\\
h_{r,s} &= \frac{(r+rp-sp)^2-1}{4p(p+1)},
 & 1\leq r \leq p-1, \ 1\leq s \leq p,
\nonumber 
\end{align}

For $c=1/2$ Ising CFT, $p=3,\ p'=4$.  $(r,s)=(1,1)$ and $(2,3)$ correspond to
the identity primary field $1$.  $(r,s)=(1,2)$ and $(2,2)$ correspond to
primary field $\si$ with $h_\si=\frac1{16}$.  $(r,s)=(1,3)$ corresponds to
primary field $\psi$ with $h_{\psi}=\frac12$.  In the basis of $\{\chi_\one,
\chi_\psi,\chi_\si\}$, the modular transformation is given by
\begin{align}
\label{IsingST}
S_\text{Is}=\frac{1}{2}
\begin{pmatrix}
1 & 1 & \sqrt{2} \\ 1 & 1 & -\sqrt{2} \\ \sqrt{2} & -\sqrt{2} & 0
\end{pmatrix}
\nonumber\\
T_\text{Is}= \ee^{-\ii \frac{\pi}{24}}
\begin{pmatrix}
1 & 0 & 0 \\ 0 & -1 & 0 \\ 0 & 0 & \ee^{\ii \frac{2\pi}{16}}
\end{pmatrix}
\end{align}

\subsection{$u1_M$ CFT}
\label{U1CFT}

$u1_M$ current algebra is generated by the current $\partial_z \varphi (z)$ and $\ee^{\ii \sqrt{M}\varphi}$. The primary fields  of the current algebra are $\ee^{\ii \frac{m}{\sqrt{M}}\varphi}, 0\leq m\leq M-1$. The character $\chi^{u1_M}_m$ of $u1_M$ CFT is given by 
\begin{align}
\label{u1chi}
 \chi^{u1_M}_m(\tau) =& \eta^{-1}(q)
\sum_{n=-\infty}^\infty q^{\frac{1}{2} (\frac{m}{R}+n R)^2} ,
\end{align}
where $0 \leq m < M$ and $R^2=M$. Under modular transformation $S$ and $T$, the
characters transform as follows,
\begin{align}
\label{u1ST}
 &\chi^{u1_M}_i(-\frac{1}{\tau}) =
 S_{ij} \chi^{u1_M}_j(\tau), &
S_{ij} &= \frac{ \ee^{-\ii 2\pi \frac{ij}{M} }}{\sqrt{M}},
\\
 &\chi^{u1_M}_i(\tau+1)= 
T_{ij} \chi^{u1_M}_j(\tau), &
T_{\ij} &=
\ee^{-\ii \frac{2\pi}{24}}\ee^{\ii 2\pi  \frac{i^2}{2M}}\del_{ij} .
\nonumber 
\end{align}

In the case of semion model, the left-moving part has two sectors, the vacuum and semion sector. They are primary fields of $u1_2$ current algebra. 
\begin{align}
\label{u12chi}
\chi^{u1_2}_0=&\eta (q)^{-1}\sum_{n\in\ZZ} q^{\frac{1}{2}(2n)2^{-1} (2n)}\nonumber=\eta (q)^{-1}\sum_{n\in \ZZ}q^{n^2}\\
\chi^{u1_2}_{1}=&\eta (q)^{-1}\sum_{n\in\ZZ}q^{(n+\frac{1}{2}) (n+\frac{1}{2})}
\end{align}

\subsection{$su2_k$ CFT}
\label{SU2CFT}

The CFT of the level-$k$ $SU(2)$ current algebra, $su2_k$, has characters
$\chi_{j}^{su2_k}(\tau)$:
\begin{align}
\label{su2chi}
\chi_{j}^{su2_k}(q)=&\frac{q^{(2j+1)^2/4(k+2)}}{[\eta (q)]^3}
\\
&\cdot \sum_{n\in\Z} \left[2j+1+2n(k+2)\right] q^{n[2j+1+(k+2)n]}
\nonumber 
\end{align}
where $j\in \left\{0,\frac{1}{2},\cdots,\frac{k}{2}\right\}$. Their modular
transformations are
\begin{align}
&\chi^{su2_k}_j(-1/\tau)=\sum_{l\in\cP}S_{jl}\chi^{su2_k}_l(\tau),\nn\\
&S_{jl}=\sqrt{\frac{2}{k+2}}\sin \left[ \frac{\pi (2j+1)(2l+1)}{k+2}\right]\\
&\chi^{su2_k}_j(\tau+1)=\ee^{-\ii \frac{2 \pi}{24}\frac{3k}{(k+2)}}\ee^{\ii 2\pi \frac{j(j+1)}{k+2}}\chi^{su2_k}_j(\tau). \nn
\end{align}

\subsection{Exceptional current algebra CFT}\label{g2chi}
Both $(G_2)_1$ and $(F_4)_1$ characters have the form as follows \cite{mathur1989reconstruction}, 
\begin{align}
\chi_0=\left[\frac{\lambda (1-\lambda)}{16}\right]^{\frac{1-x}{6}} {}_2F_1\left(\frac{1}{2}-\frac{1}{6}x, \frac{1}{2}-\frac{1}{2}x ; 1-\frac{1}{3}x ;\lambda \right)\nn\\
\chi_1=N\left[\frac{\lambda (1-\lambda)}{16}\right]^{\frac{1+x}{6}} {}_2F_1\left(\frac{1}{2}+\frac{1}{6}x, \frac{1}{2}+\frac{1}{2}x;  1+\frac{1}{3}x ;\lambda \right)
\end{align}
where $\lambda (\tau)=\left(\frac{\theta_2(\tau)}{\theta_3(\tau)}\right)^4$, in terms of theta functions,
\begin{align}
\theta_2(\tau)=&\sum_{n\in\ZZ} q^{\frac{1}{2}\left(n+\frac{1}{2}\right)^2},\,~~\theta_4(\tau)=&\sum_{n\in\ZZ} (-1)^nq^{\frac{n^2}{2}}
\end{align}
 Under modular transformation, $T:\lambda\rightarrow \lambda (\lambda-1)$, and $S: \lambda\rightarrow 1-\lambda$. 
\begin{align}
{}_2F_1 (a,b;c;z)=&\sum_{n=0}^\infty \frac{(a)_n (b)_n}{(c)_n}\frac{z^n}{n!}\\
(q)_n=&\begin{cases}
1 & n=0 \\
\frac{(q+n-1)!}{(q-1)!} & n>0
\end{cases}
\end{align}
is the hypergeometric function defined for $|z|<1$, and $x=1+\frac{c}{2}$. The parameters for some examples are
\begin{align}
(G_2)_1:& \quad N=7,\; x=\frac{12}{5},\nn\\
(F_4)_1:&\quad N=26,\; x=\frac{18}{5}\\
(E_8)_1:&\quad N=2,\; x=4.\nn
\end{align}

\section{Non-on-site $Z_2$ symmetry transformations}
\label{Z2trns}

The first non-on-site $Z_2$ symmetry transformation
\eq{XCZ} transforms $\si^x_i$ in the following way (see \eqn{CZij})
\begin{align}
&\ \ \ \  
(\prod_j \si^x_j \prod_j CZ_{j,j+1} )
\si^x_i
(\prod_j \si^x_j \prod_j CZ_{j,j+1} )
\nonumber\\
&=
\frac { 1 +\si^z_{i-1} +\si^z_i -\si^z_{i-1} \si^z_i}2
\frac {1 +\si^z_i +\si^z_{i+1} -\si^z_i \si^z_{i+1}}2 
\si^x_i 
\nonumber\\
&\ \ \ \
\frac { 1 +\si^z_{i-1} +\si^z_i -\si^z_{i-1} \si^z_i}2
\frac {1 +\si^z_i +\si^z_{i+1} -\si^z_i \si^z_{i+1}}2 
\nonumber\\ 
& = 
\frac { 1 +\si^z_{i-1} +\si^z_i -\si^z_{i-1} \si^z_i}2
\frac {1 +\si^z_i +\si^z_{i+1} -\si^z_i \si^z_{i+1}}2 
\nonumber\\
&\ \ \ \
\frac { 1 +\si^z_{i-1} -\si^z_i +\si^z_{i-1} \si^z_i}2
\frac {1 -\si^z_i +\si^z_{i+1} +\si^z_i \si^z_{i+1}}2 
\si^x_i 
\nonumber\\ 
& = 
\Big(\frac { 1 +\si^z_{i-1} +\si^z_i -\si^z_{i-1} \si^z_i}2
\frac { 1 +\si^z_{i-1} -\si^z_i +\si^z_{i-1} \si^z_i}2 \Big)
\nonumber\\
&\ \ \ \
\Big(\frac {1 +\si^z_i +\si^z_{i+1} -\si^z_i \si^z_{i+1}}2 
\frac {1 -\si^z_i +\si^z_{i+1} +\si^z_i \si^z_{i+1}}2 \Big)
\si^x_i 
\nonumber\\ 
& = 
\frac { (1 +\si^z_{i-1}) -(1 -\si^z_{i-1}) }2
\frac { (1 +\si^z_{i+1}) -(1 - \si^z_{i+1})}2 
\si^x_i 
\nonumber\\ 
& = 
\si^z_{i-1}
\si^x_i 
\si^z_{i+1}
\end{align}
The second non-on-site $Z_2$ symmetry transformation
\eq{Xs} transforms $\si^x_i$ in the same way (see \eqn{sij}):
\begin{align}
&\ \ \ \  
(\prod_j \si^x_j \prod_j s_{j,j+1} )
\si^x_i
(\prod_j \si^x_j \prod_j s_{j,j+1} )
\nonumber\\
&=
\frac { 1 -\si^z_{i-1} +\si^z_i +\si^z_{i-1} \si^z_i}2
\frac {1 -\si^z_i +\si^z_{i+1} +\si^z_i \si^z_{i+1}}2 
\si^x_i 
\nonumber\\ 
&\ \ \ \
\frac { 1 -\si^z_{i-1} +\si^z_i +\si^z_{i-1} \si^z_i}2
\frac {1 -\si^z_i +\si^z_{i+1} +\si^z_i \si^z_{i+1}}2 
\nonumber\\
& = 
\frac { 1 -\si^z_{i-1} +\si^z_i +\si^z_{i-1} \si^z_i}2
\frac {1 -\si^z_i +\si^z_{i+1} +\si^z_i \si^z_{i+1}}2 
\nonumber\\
&\ \ \ \
\frac { 1 -\si^z_{i-1} -\si^z_i -\si^z_{i-1} \si^z_i}2
\frac {1 +\si^z_i +\si^z_{i+1} -\si^z_i \si^z_{i+1}}2 
\si^x_i 
\nonumber\\ 
& = 
\Big(\frac { 1 -\si^z_{i-1} +\si^z_i +\si^z_{i-1} \si^z_i}2
\frac { 1 -\si^z_{i-1} -\si^z_i -\si^z_{i-1} \si^z_i}2 \Big)
\nonumber\\
&\ \ \ \
\Big(\frac {1 -\si^z_i +\si^z_{i+1} +\si^z_i \si^z_{i+1}}2 
\frac {1 +\si^z_i +\si^z_{i+1} -\si^z_i \si^z_{i+1}}2 \Big)
\si^x_i 
\nonumber\\ 
& = 
\frac { (1 -\si^z_{i-1}) -(1 +\si^z_{i-1}) }2
\frac { (1 +\si^z_{i+1}) -(1 - \si^z_{i+1})}2 
\si^x_i 
\nonumber\\ 
& = 
\si^z_{i-1}
\si^x_i 
\si^z_{i+1}
\end{align}

\section{Topological path integral on a space-time with world
lines domain walls}
\label{path}

\begin{figure}[tb]
\begin{center}
\includegraphics[scale=0.6]{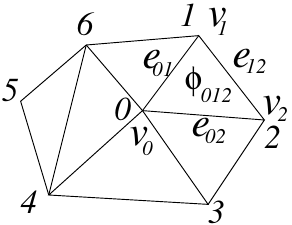} \end{center}
%Fig. 1
\caption{ 
A 2-dimensional complex.  The vertices (0-simplices) are labeled by $i$.  The
edges (1-simplices) are labeled by $\<ij\>$.  The faces (2-simplices) are
labeled by $\<ijk\>$.  The degrees of freedoms may live on the vertices
(labeled by $v_i$), on the edges (labeled by $e_{ij}$) and on the faces
(labeled by $\phi_{ijk}$).
}
\label{comp}
\end{figure}
\begin{figure}[tb]
\begin{center}
\includegraphics[scale=0.6]{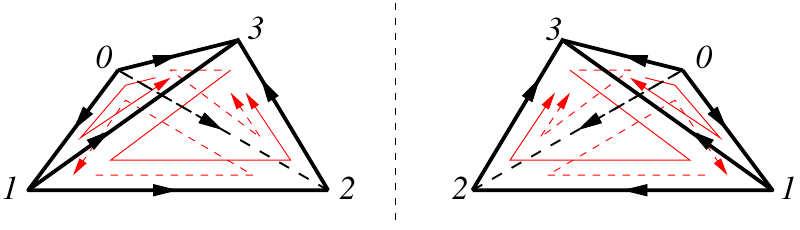} \end{center}
%Fig. 2
\caption{ (Color online) Two branched simplices with opposite orientations.
(a) A branched simplex with positive orientation and (b) a branched simplex
with negative orientation.  }
\label{mir}
\end{figure}

\subsection{Space-time lattice and branching structure}

To find the conditions on the domain-wall data, we need to use extensively the
space-time path integral.  So we will first describe how to define a space-time
path integral. We first triangulate the $3$-dimensional space-time to obtain a
simplicial complex $\cM^3$ (see Fig. \ref{comp}).  Here we assume that all
simplicial complexes are of bounded geometry in the sense that the number of
edges that connect to one vertex is bounded by a fixed value.  Also, the number
of triangles that connect to one edge is bounded by a fixed value, \etc.

In order to define a generic lattice theory on the space-time complex $\cM^3$,
it is important to give the vertices of each simplex a local order.  A nice
local scheme to order  the vertices is given by a branching
structure.\cite{C0527,CGL1314} A branching structure is a choice of
the orientation of each edge in the $n$-dimensional complex so that there is no
oriented loop on any triangle (see Fig. \ref{mir}).

The branching structure induces a \emph{local order} of the vertices on each
simplex.  The first vertex of a simplex is the vertex with no incoming edges,
and the second vertex is the vertex with only one incoming edge, \etc.  So the
simplex in  Fig. \ref{mir}a has the following vertex ordering: $0<1<2<3$.

The branching structure also gives the simplex (and its sub simplexes) an
orientation denoted by $s_{ij \cdots k}=1,*$.  Fig. \ref{mir} illustrates two
$3$-simplices with opposite orientations $s_{0123}=1$ and $s_{0123}=*$.  The
red arrows indicate the orientations of the $2$-simplices which are the
subsimplices of the $3$-simplices.  The black arrows on the edges indicate the
orientations of the $1$-simplices.

The degrees of freedom of our lattice model live on the vertices  (denoted by
$v_i$ where $i$ labels the vertices), on the edges (denoted by $e_{ij}$ where
$\<ij\>$ labels the edges), and on other high dimensional simplicies of the
space-time complex (see Fig. \ref{comp}).

\begin{figure}[t]
\begin{center}
\includegraphics[scale=0.6]{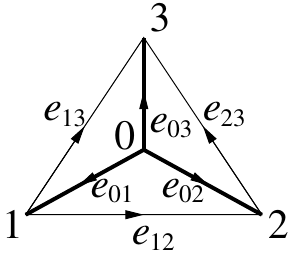}
%Fig. 5
\end{center}
\caption{
The tensor $\tC{C}0123$ is associated with a tetrahedron, which has a branching
structure.  If the vertex-0 is above the triangle-123, then the tetrahedron
will have an orientation $s_{0123}=*$.  If the vertex-0 is below the
triangle-123, the tetrahedron will have an orientation $s_{0123}=1$. The
branching structure gives the vertices a local order: the $i^{th}$ vertex has
$i$ incoming edges.  
}
\label{tetr}
\end{figure}

\subsection{Discrete path integral}

In this paper, we will only consider a type of 2+1D path integral that can be
constructed from a tensor set $T$ of two real and one complex tensors:
$T=(w_{v_0}, \tAw{d}01,\tC{C}0123)$.  The complex tensor $\tC{C}0123$ can be
associated with a tetrahedron, which has a branching structure (see Fig.
\ref{tetr}).  A branching structure is a choice of an orientation of each edge in
the complex so that there is no oriented loop on any triangle (see Fig.
\ref{tetr}).  Here the $v_0$ index is associated with the vertex-0, the
$e_{01}$ index is associated with the edge-$01$, and the $\phi_{012}$ index is
associated with the triangle-$012$.  They represent the degrees of freedom on
the vertices, edges, and triangles.

Using the tensors, we can define the path integral on any 3-complex
that has no boundary:
\begin{align}
\label{Z3d}
 Z(\cM^3)&=\sum_{ v_0,\cdots; e_{01},\cdots; \phi_{012},\cdots}
\prod_\text{vertex} w_{v_{0}} 
\prod_\text{edge} \tAw{d}01\times
\\
&\ \ \ \ \ \ \ \ \ \ 
\prod_\text{tetra} [\tC{C}0123 ]^{s_{0123}}
\nonumber 
\end{align}
where $\sum_{v_0,\cdots; e_{01},\cdots; \phi_{012},\cdots}$ sums over all the
vertex indices, the edge indices, and the face indices, $s_{0123}=1$ or $*$
depending on the orientation of tetrahedron (see Fig.  \ref{tetr}).  We believe
such type of path integral can realize any 2+1D topological order.

\subsection{Path integral on space-time with natural boundary}
\label{Nboundary}

On the complex $\cM^3$ with boundary: $\cB^2= \prt \cM^3$, the partition
function is defined differently:
\begin{align}
\label{Z3dB}
 Z(\cM^3) & =\sum_{ \{ v_i; e_{ij}; \phi_{ijk} \} }
\prod_{\text{vertex}\notin \cB^2} w_{v_{0}} 
\prod_{\text{edge}\notin \cB^2} \tAw{d}01\times
\\
&\ \ \ \ \ \ \ \ \ \ 
\prod_\text{tetra} [\tC{C}0123 ]^{s_{0123}}
\nonumber 
\end{align}
where $\sum_{v_i; e_{ij}; \phi_{ijk}}$ only sums over the vertex indices, the
edge indices, and the face indices that are not on the boundary.  The resulting
$Z(\cM^3)$ is actually a complex function of $v_{i}$'s, $e_{ij}$'s, and
$\phi_{ijk}$'s on the boundary $\cB^2$: $Z(\cM^3;\{v_{i};e_{ij};\phi_{ijk}\})$.
Such a function is a vector in the vector space $\cV_{\cB^2}$.  (The vector
space $\cV_{\cB^2}$ is the space of all complex function of the boundary
indices on the boundary complex $\cB^2$: $\Psi(\{v_{i};e_{ij};\phi_{ijk}\})$.)
We will denote such a vector as $|\Psi(\cM^3)\>$.
boundary) are attached with the tensors $w_{v_{i}}$ and $\tAw{d}01$.  
The boundary \eq{Z3dB} defined above is called a natural boundary
of the path integral.

We also note that only the vertices and the edges in the bulk (\ie not on the boundaries.) When we glue two boundaries together, those tensors $w_{v_{i}}$ and
$\tAw{d}ij$ are added back. For example, let $\cM^3$ and $\cN^3$ to have the
same boundary (with opposite orientations)
\begin{align}
 \prt \cM^3 = -  \prt \cN^3 = \cB^2
\end{align}
which give rise to wavefunctions on the boundary $|\Psi(\cM^3)\>$ and
$\<\Psi(\cN^3)|$ after the path integral in the bulk.  Gluing two boundaries
together corresponds to the inner product $\<\Psi(\cN^3)|\Psi(\cM^3)\>$.  So the
tensors $w_{v_{i}}$ and $\tAw{d}ij$ defines the inner product in the boundary
Hilbert space $\cV_{\cB^2}$.  Therefore, we require $w_{v_{i}}$ and $\tAw{d}ij$
to satisfy the following unitary condition
\begin{align}
 w_{v_{i}} > 0, \ \ \ \tAw{d}ij >0.
\end{align}

\subsection{Topological path integral}
\label{toppath}

We notice that the above path integral is defined for any space-time lattice.
The partition function $Z(\cM^3)$ depends on the choices of the space-time
lattice.  For example, $Z(\cM^3)$ depends on the number of the cells in
space-time, which give rise to the leading volume dependent term,
in the large space-time limit (\ie the thermodynamic limit)
\begin{align}
 Z(\cM^3) = \ee^{-\eps V} Z^\text{top}(\cM^3)
\end{align}
where $V$ is the space-time volume, $\eps$ is the energy density of the
ground state, and $Z^\text{top}(\cM^3)$ is the volume independent partition
function. It was conjectured that the volume independent partition function
$Z^\text{top}(\cM^3)$ in the thermodynamic limit, as a function of closed
space-time $\cM^3$, is a topological invariant that can fully characterize
topological order.\cite{KW1458,WW180109938} So it is very desirable to fine
tune the path integral to make the energy density $\eps=0$.  This can be
achieved by fine-tuning the tensors $w_{v_{i}}$ and $\tAw{d}ij$.  But we can do
better.  We can choose the tensor $(w_{v_0}$, $\tAw{d}01$, $\tC{C}0123)$
to be the fixed-point tensor-set under the renormalization group flow of the
tensor network.\cite{LN0701,GW0931} In this case, not only the volume factor
$\ee^{-\eps V}$ disappears, the volume independent partition function
$Z^\text{top}(\cM^3)$ is also  re-triangulation invariant, for any size of
space-time lattice. In this case, we refer the path integral as a topological
path integral, and denote the resulting partition function as
$Z^\text{top}(\cM^3)$.  $Z^\text{top}$ is also referred as the volume
independent the partition function, which is a very important concept, since
only the volume independent partition functions correspond to topological
invariants.  In particular, it was conjectured that such kind of  topological
path integrals describes all the topological order with gappable boundary.  For
details, see \Ref{KW1458,WW180109938}.  

\begin{figure}[t]
\begin{center}
\includegraphics[scale=0.5]{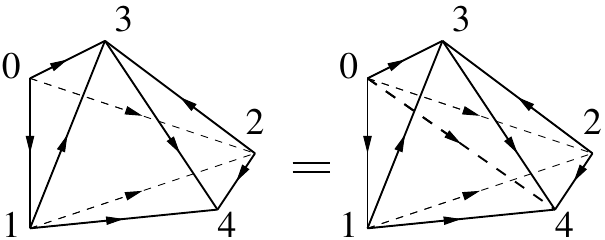}
%Fig. 6
\end{center}
\caption{
A re-triangulation of a 3D complex.
}
\label{2to3}
\end{figure}
\begin{figure}[t]
\begin{center}
\includegraphics[scale=0.5]{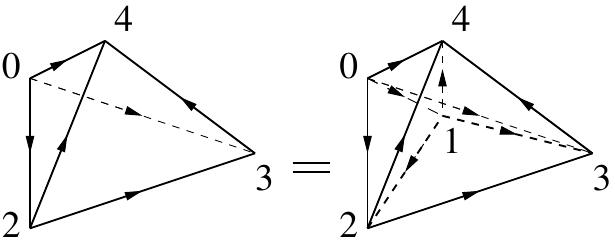}
%Fig. 7
\end{center}
\caption{
A re-triangulation of another 3D complex.
}
\label{1to4}
\end{figure}

The invariance of partition function $Z$ under the re-triangulation in Fig.
\ref{2to3} and \ref{1to4} requires that
\begin{align}
\label{CC23}
&\ \ \
\sum_{\phi_{123}} \tC{C}0123 \tC{C}1234
\nonumber\\
&=
\sum_{e_{04}} \tAw{d}04
\sum_{ \phi_{014} \phi_{024} \phi_{034} }
\tC{C}0124 
\\
&\ \ \ \ \ \ \
\tC{C^*}0134
\tC{C}0234 .
\nonumber 
\end{align}
\begin{align}
\label{CC14}
& \ \ \ \
\tC{C}0234
\\
&=
\sum_{e_{01}e_{12}e_{13}e_{14},v_1} w_{v_1} 
\tAw{d}01 \tAw{d}12 \tAw{d}13 \tAw{d}14 
\hskip -10mm
\sum_{ 
\phi_{012} 
\phi_{013} 
\phi_{014} 
\phi_{123} 
\phi_{124} 
\phi_{134} 
}
\nonumber\\
&\ \ \ \ \ 
\tC{C}0123
\tC{C^*}0124 
\nonumber\\
&\ \ \ \ \ 
\tC{C}0134
\tC{C}1234
\nonumber 
\end{align}
We would like to mention that there are other similar conditions for different
choices of the branching structures.  The branching structure of a tetrahedron
affects the labeling of the vertices. For more details, see \Ref{ZW180809394}.

\subsection{Topological path integral with world-lines}
\label{toppathW}

In this paper, we also need to use the space-time path integral with
world-lines of topological excitations.  We denote the resulting partition
function as
\begin{align}
\label{Zlines}
Z \bpm \includegraphics[scale=.50]{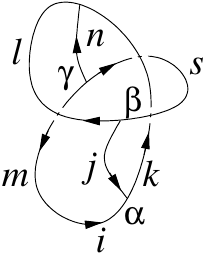} \epm ,
\end{align}
where $i,j,k,\cdots \in \{1,2,\cdots,N\}$ label the type of topological
excitations, and $\al,\bt,\ga$ label the different fusion channels (\ie
different choices of actions at the junction of three world-lines).  The world
lines are defined via a different choice of tensors for simplexes that touch
the world-lines.  In this paper, we will choose the tensors very
carefully, so that the path integral with world-lines is also re-triangulation
invariant (even for the  re-triangulations that involve the world-lines).  The
different choices of re-triangulation-invariant world-lines are labeled by the
different types of topological excitations.  In this paper, we will only
consider those topological path integrals with re-triangulation invariance.

\section{Renormalization group flow of marginal perturbations of $SU(2)_1$ CFT}\label{RG}

\begin{figure}[tb]
\centering
\begin{tikzpicture}[baseline={(current bounding box.center)}]
\tikzset{->-/.style={decoration={
  markings,
  mark=at position 1. with {\arrow{>}}},postaction={decorate}}}
\draw (0,0) rectangle (2,2); \draw (0.4,0.6) rectangle (2.4,2.6);
\draw (0,0) -- (0.4,0.6);\draw (2,0) -- (2.4,0.6);\draw (0,2)--(0.4,2.6);\draw (2,2)--(2.4,2.6);
\draw [draw=blue, thick,->,>=stealth] (0,0)--(0.5,13/12*0.5);
\draw [draw=blue, thick,->,>=stealth] (0.5,13/12*0.5)--(2.1,13/12*2.1);
\draw [draw=blue, thick] (2.1,13/12*2.1)--(2.4,2.6);
\draw [draw=blue, thick,->,>=stealth] (2,2)--(1.6,1.2*7/8+0.6);
\draw [draw=blue, thick,->,>=stealth] (2,2)--(0.7,0.3*7/8+0.6);
\draw [draw=blue, thick] (0.7,0.3*7/8+0.6)--(0.4,0.6);
\draw [draw=blue, thick,->,>=stealth] (0.4,2.6)--(0.7,2.6-13/8*0.3);
\draw [draw=blue, thick,->,>=stealth] (0.7,2.6-13/8*0.3)--(1.8,2.6-13/8*1.4);
\draw [draw=blue, thick] (1.8,2.6-13/8*1.4)--(2,0);
\draw [draw=blue, thick,->,>=stealth] (2.4,0.6)--(1.8,0.6+7/12*0.6);
\draw [draw=blue, thick,->,>=stealth] (1.8,0.6+7/12*0.6)--(0.5,0.6+7/12*1.9);
\draw [draw=blue, thick] (0.5,0.6+7/12*1.9)--(0,2);
\node at (-0.8,-0.3) {$(---)$};\node at (2.4,-0.3) {$(+--)$};\node at (3.2,0.6) {$(++-)$};\node at (3.2,2) {$(+-+)$};\node at (3.2,2.6) {$(+++)$};\node at (-0.4,2.6) {$(-++)$};\node at (-0.8,2) {$(--+)$};\node at (-0.8,0.6) {$(-+-)$};
\draw [->,>=stealth] (-2.7,0.2)--(-1.8,0.2);\draw [->,>=stealth] (-2.7,0.2)--(-2.3,0.8);\draw [->,>=stealth] (-2.7,0.2)--(-2.7,1.1);
\node at (-1.8,0) {$g_1$};\node at (-2.05,0.8) {$g_2$};\node at (-2.95,1.1) {$g_3$};
\draw[->-,draw=red,thick] (1.2-0.1,1.3+0.31) .. controls (1.2,1.36) .. (1.2+0.2,1.3+0.36);
\draw[->-,draw=red,thick] (1.2-0.2,1.3-0.31) .. controls (1.2,1.27) .. (1.2+0.1,1.3-0.36);
\draw[->-,draw=red,thick] (1.2-0.5,1.3+0.63) .. controls (1.2-0.1,1.3+0.1) .. (1.2-0.5,1.3+0.4);
\draw[->-,draw=red,thick] (1.2+0.5,1.3-0.4) .. controls (1.2+0.1,1.3-0.1) .. (1.2+0.5,1.3-0.63);
\end{tikzpicture}
\caption{The RG flow of $\delta S=\sum_{i=1,2,3}\int d^2xg_i J_i\bar J_i$. There are only fixed lines, shown as the blue lines. There are no stable sheets or regions, as indicated by the orange flow arrows. The corners are labeled by $(s_1,s_2,s_3)$.}
\label{rg}
\end{figure}
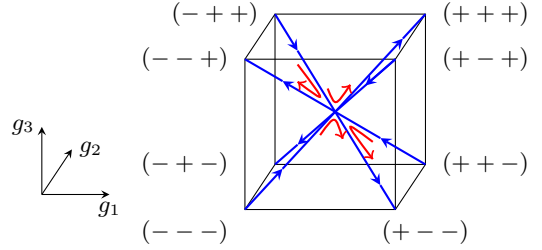
  
There are in total $9$ terms of marginal perturbations in $SU(2)_1$ CFT,
composed of left and right currents. Let us first consider the following three
couplings: 
\begin{align}
  S_{\text{int}}=\sum_{i=1}^3 \int g^{i} O_{i}, \ \ \ \ \ \ 
O_{i}=J_i\bar J_i.  
\end{align}
The renormalization group (RG) equations have the form
\begin{align}
\dot{g}_i=\alpha_{ijk}g_jg_k,
\end{align}
where $\alpha_{ijk}$ is proportional to the operator product expansion,
\begin{align}
\alpha_{ijk}=\langle O_i O_j O_k\rangle= \langle J_i J_j J_k\rangle\langle \bar J_i \bar J_j \bar J_k\rangle=(\epsilon_{ijk})^2.
\end{align}
It follows that 
\begin{align}
\dot{g}_1= g_2g_3,\quad \dot g_2=g_3g_1,\quad \dot g_3=g_1g_2
\label{3perb}
\end{align}

The solution of the beta function has $4$ fixed lines. To solve them,
take the form $g_i(t)=\lambda_i f(t)$, and one finds
$\frac{\lambda_1\lambda_2}{\lambda_3}=\frac{\lambda_2\lambda_3}{\lambda_1}=\frac{\lambda_3\lambda_1}{\lambda_2}$.
Therefore, $\lambda_i=s_i\alpha$, where $\alpha \geq 0$, $s_i=\pm 1$ to be
determined.  The RG equations become
\begin{align}
\dot f(t)= s\alpha f^2(t)
\end{align}
where $s=s_1s_2s_3$. The solution is
\begin{align}
f(t)=\frac{f(0)}{1-s\alpha f(0) t}
\end{align}
And this leads to the RG solution of fixed lines
\begin{align}
g_i(t)=\frac{g_i (0)}{1-s|g_i(0)|t},\quad |g_1(0)|=|g_2(0)|=|g_3(0)|
\end{align}
%Where $s>0$, there is always a finite positive $t_c=\frac{1}{|g_i(0)|}$ at
%which $g_i(t_c)$'s diverge. 
We find that
\begin{itemize}
\item when $s>0$, the following four fixed lines flow towards infinity
\begin{align}
g_1(t)&=g_2(t)=g_3(t)>0,
\nonumber\\
 g_1(t)&=-g_2(t)=-g_3(t)>0
\nonumber\\
 -g_1(t)&=g_2(t)=-g_3(t)>0,
\nonumber\\
 -g_1(t)&=-g_2(t)=g_3(t)>0.
\end{align}

\item when $s<0$, the following four fixed lines flow towards $g_1=g_2=g_3=0$
\begin{align}
g_1(t) &= g_2(t)=g_3(t)<0,
\nonumber\\
 g_1(t) &=-g_2(t)=-g_3(t)<0
\nonumber\\
 -g_1(t)&=g_2(t)=-g_3(t)<0,
\nonumber\\
 -g_1(t) &=-g_2(t)=g_3(t)<0.
\end{align}
\end{itemize}
This allows us to show that there are no stable regions or sheets in the
$(g_1,g_2,g_3)$ parameter space, as illustrated in Fig. \ref{rg}.

Through the above example, we see a general pattern.  If there is only one
marginally relevant coupling, \ie if we are on a fixed line, then there is a
finite region, such that all the couplings in that region flow to zero.  This
finite region represents the  region of stable gapless phase.  If there are two
marginally relevant couplings, \ie if we are on a plane spanned by two fixed
lines, then there is no finite region where the couplings flow to zero.  When
there are more marginally relevant couplings, the system is getting even more
unstable.  So we believe that, for our case with 9  marginally relevant
couplings, the corresponding CFT is unstable.

\bibliography{../../../bib/all,../../../bib/publst,../local}

\begin{thebibliography}{56}%
\makeatletter
\providecommand \@ifxundefined [1]{%
 \@ifx{#1\undefined}
}%
\providecommand \@ifnum [1]{%
 \ifnum #1\expandafter \@firstoftwo
 \else \expandafter \@secondoftwo
 \fi
}%
\providecommand \@ifx [1]{%
 \ifx #1\expandafter \@firstoftwo
 \else \expandafter \@secondoftwo
 \fi
}%
\providecommand \natexlab [1]{#1}%
\providecommand \enquote  [1]{``#1''}%
\providecommand \bibnamefont  [1]{#1}%
\providecommand \bibfnamefont [1]{#1}%
\providecommand \citenamefont [1]{#1}%
\providecommand \href@noop [0]{\@secondoftwo}%
\providecommand \href [0]{\begingroup \@sanitize@url \@href}%
\providecommand \@href[1]{\@@startlink{#1}\@@href}%
\providecommand \@@href[1]{\endgroup#1\@@endlink}%
\providecommand \@sanitize@url [0]{\catcode `\\12\catcode `\$12\catcode
  `\&12\catcode `\#12\catcode `\^12\catcode `\_12\catcode `\%12\relax}%
\providecommand \@@startlink[1]{}%
\providecommand \@@endlink[0]{}%
\providecommand \url  [0]{\begingroup\@sanitize@url \@url }%
\providecommand \@url [1]{\endgroup\@href {#1}{\urlprefix }}%
\providecommand \urlprefix  [0]{URL }%
\providecommand \Eprint [0]{\href }%
\providecommand \doibase [0]{http://dx.doi.org/}%
\providecommand \selectlanguage [0]{\@gobble}%
\providecommand \bibinfo  [0]{\@secondoftwo}%
\providecommand \bibfield  [0]{\@secondoftwo}%
\providecommand \translation [1]{[#1]}%
\providecommand \BibitemOpen [0]{}%
\providecommand \bibitemStop [0]{}%
\providecommand \bibitemNoStop [0]{.\EOS\space}%
\providecommand \EOS [0]{\spacefactor3000\relax}%
\providecommand \BibitemShut  [1]{\csname bibitem#1\endcsname}%
\let\auto@bib@innerbib\@empty
%</preamble>
\bibitem [{\citenamefont {Adler}(1969)}]{A6926}%
  \BibitemOpen
  \bibfield  {author} {\bibinfo {author} {\bibfnamefont {S.~L.}\ \bibnamefont
  {Adler}},\ }\href {\doibase 10.1103/physrev.177.2426} {\bibfield  {journal}
  {\bibinfo  {journal} {Phys. Rev.}\ }\textbf {\bibinfo {volume} {177}},\
  \bibinfo {pages} {2426} (\bibinfo {year} {1969})}\BibitemShut {NoStop}%
\bibitem [{\citenamefont {Bell}\ and\ \citenamefont {Jackiw}(1969)}]{BJ6947}%
  \BibitemOpen
  \bibfield  {author} {\bibinfo {author} {\bibfnamefont {J.}~\bibnamefont
  {Bell}}\ and\ \bibinfo {author} {\bibfnamefont {R.}~\bibnamefont {Jackiw}},\
  }\href@noop {} {\bibfield  {journal} {\bibinfo  {journal} {Nuovo Cimento}\
  }\textbf {\bibinfo {volume} {60A}},\ \bibinfo {pages} {47} (\bibinfo {year}
  {1969})}\BibitemShut {NoStop}%
\bibitem [{\citenamefont {Witten}(1985)}]{W8597}%
  \BibitemOpen
  \bibfield  {author} {\bibinfo {author} {\bibfnamefont {E.}~\bibnamefont
  {Witten}},\ }\href {\doibase 10.1007/bf01212448} {\bibfield  {journal}
  {\bibinfo  {journal} {Commun. Math. Phys.}\ }\textbf {\bibinfo {volume}
  {100}},\ \bibinfo {pages} {197} (\bibinfo {year} {1985})}\BibitemShut
  {NoStop}%
\bibitem [{\citenamefont {’t Hooft}(1980)}]{H8035}%
  \BibitemOpen
  \bibfield  {author} {\bibinfo {author} {\bibfnamefont {G.}~\bibnamefont {’t
  Hooft}},\ }\href@noop {} {\bibfield  {journal} {\bibinfo  {journal} {NATO
  Adv. Study Inst. Ser. B Phys.}\ }\textbf {\bibinfo {volume} {59}},\ \bibinfo
  {pages} {135} (\bibinfo {year} {1980})}\BibitemShut {NoStop}%
\bibitem [{\citenamefont {Callan}\ and\ \citenamefont {Harvey}(1985)}]{CH8527}%
  \BibitemOpen
  \bibfield  {author} {\bibinfo {author} {\bibfnamefont {C.}~\bibnamefont
  {Callan}}\ and\ \bibinfo {author} {\bibfnamefont {J.}~\bibnamefont
  {Harvey}},\ }\href {\doibase 10.1016/0550-3213(85)90489-4} {\bibfield
  {journal} {\bibinfo  {journal} {Nucl. Phys. B}\ }\textbf {\bibinfo {volume}
  {250}},\ \bibinfo {pages} {427} (\bibinfo {year} {1985})}\BibitemShut
  {NoStop}%
\bibitem [{\citenamefont {Ryu}\ \emph {et~al.}(2012)\citenamefont {Ryu},
  \citenamefont {Moore},\ and\ \citenamefont {Ludwig}}]{RML1204}%
  \BibitemOpen
  \bibfield  {author} {\bibinfo {author} {\bibfnamefont {S.}~\bibnamefont
  {Ryu}}, \bibinfo {author} {\bibfnamefont {J.~E.}\ \bibnamefont {Moore}}, \
  and\ \bibinfo {author} {\bibfnamefont {A.~W.~W.}\ \bibnamefont {Ludwig}},\
  }\href {\doibase 10.1103/physrevb.85.045104} {\bibfield  {journal} {\bibinfo
  {journal} {Phys. Rev. B}\ }\textbf {\bibinfo {volume} {85}},\ \bibinfo
  {pages} {045104} (\bibinfo {year} {2012})},\ \Eprint
  {http://arxiv.org/abs/arXiv:1010.0936} {arXiv:1010.0936} \BibitemShut
  {NoStop}%
\bibitem [{\citenamefont {Wen}(2013)}]{W1313}%
  \BibitemOpen
  \bibfield  {author} {\bibinfo {author} {\bibfnamefont {X.-G.}\ \bibnamefont
  {Wen}},\ }\href {\doibase 10.1103/physrevd.88.045013} {\bibfield  {journal}
  {\bibinfo  {journal} {Phys. Rev. D}\ }\textbf {\bibinfo {volume} {88}},\
  \bibinfo {pages} {045013} (\bibinfo {year} {2013})},\ \Eprint
  {http://arxiv.org/abs/arXiv:1303.1803} {arXiv:1303.1803} \BibitemShut
  {NoStop}%
\bibitem [{\citenamefont {Kong}\ and\ \citenamefont {Wen}(2014)}]{KW1458}%
  \BibitemOpen
  \bibfield  {author} {\bibinfo {author} {\bibfnamefont {L.}~\bibnamefont
  {Kong}}\ and\ \bibinfo {author} {\bibfnamefont {X.-G.}\ \bibnamefont {Wen}},\
  }\href@noop {} {\  (\bibinfo {year} {2014})},\ \Eprint
  {http://arxiv.org/abs/arXiv:1405.5858} {arXiv:1405.5858} \BibitemShut
  {NoStop}%
\bibitem [{\citenamefont {{Kong}}\ \emph {et~al.}(2015)\citenamefont {{Kong}},
  \citenamefont {{Wen}},\ and\ \citenamefont {{Zheng}}}]{KZ150201690}%
  \BibitemOpen
  \bibfield  {author} {\bibinfo {author} {\bibfnamefont {L.}~\bibnamefont
  {{Kong}}}, \bibinfo {author} {\bibfnamefont {X.-G.}\ \bibnamefont {{Wen}}}, \
  and\ \bibinfo {author} {\bibfnamefont {H.}~\bibnamefont {{Zheng}}},\
  }\href@noop {} {\  (\bibinfo {year} {2015})},\ \Eprint
  {http://arxiv.org/abs/1502.01690} {arXiv:1502.01690} \BibitemShut {NoStop}%
\bibitem [{\citenamefont {Wen}(1989)}]{W8987}%
  \BibitemOpen
  \bibfield  {author} {\bibinfo {author} {\bibfnamefont {X.-G.}\ \bibnamefont
  {Wen}},\ }\href {\doibase 10.1103/physrevb.40.7387} {\bibfield  {journal}
  {\bibinfo  {journal} {Phys. Rev. B}\ }\textbf {\bibinfo {volume} {40}},\
  \bibinfo {pages} {7387} (\bibinfo {year} {1989})}\BibitemShut {NoStop}%
\bibitem [{\citenamefont {Wen}(1990)}]{W9039}%
  \BibitemOpen
  \bibfield  {author} {\bibinfo {author} {\bibfnamefont {X.-G.}\ \bibnamefont
  {Wen}},\ }\href {\doibase 10.1142/s0217979290000139} {\bibfield  {journal}
  {\bibinfo  {journal} {Int. J. Mod. Phys. B}\ }\textbf {\bibinfo {volume}
  {04}},\ \bibinfo {pages} {239} (\bibinfo {year} {1990})}\BibitemShut
  {NoStop}%
\bibitem [{\citenamefont {Gu}\ and\ \citenamefont {Wen}(2009)}]{GW0931}%
  \BibitemOpen
  \bibfield  {author} {\bibinfo {author} {\bibfnamefont {Z.-C.}\ \bibnamefont
  {Gu}}\ and\ \bibinfo {author} {\bibfnamefont {X.-G.}\ \bibnamefont {Wen}},\
  }\href@noop {} {\bibfield  {journal} {\bibinfo  {journal} {Phys. Rev. B}\
  }\textbf {\bibinfo {volume} {80}},\ \bibinfo {pages} {155131} (\bibinfo
  {year} {2009})},\ \Eprint {http://arxiv.org/abs/arXiv:0903.1069}
  {arXiv:0903.1069} \BibitemShut {NoStop}%
\bibitem [{\citenamefont {Chen}\ \emph {et~al.}(2011)\citenamefont {Chen},
  \citenamefont {Liu},\ and\ \citenamefont {Wen}}]{CLW1141}%
  \BibitemOpen
  \bibfield  {author} {\bibinfo {author} {\bibfnamefont {X.}~\bibnamefont
  {Chen}}, \bibinfo {author} {\bibfnamefont {Z.-X.}\ \bibnamefont {Liu}}, \
  and\ \bibinfo {author} {\bibfnamefont {X.-G.}\ \bibnamefont {Wen}},\ }\href
  {\doibase 10.1103/physrevb.84.235141} {\bibfield  {journal} {\bibinfo
  {journal} {Phys. Rev. B}\ }\textbf {\bibinfo {volume} {84}},\ \bibinfo
  {pages} {235141} (\bibinfo {year} {2011})},\ \Eprint
  {http://arxiv.org/abs/arXiv:1106.4752} {arXiv:1106.4752} \BibitemShut
  {NoStop}%
\bibitem [{\citenamefont {Chen}\ \emph {et~al.}(2013)\citenamefont {Chen},
  \citenamefont {Gu}, \citenamefont {Liu},\ and\ \citenamefont
  {Wen}}]{CGL1314}%
  \BibitemOpen
  \bibfield  {author} {\bibinfo {author} {\bibfnamefont {X.}~\bibnamefont
  {Chen}}, \bibinfo {author} {\bibfnamefont {Z.-C.}\ \bibnamefont {Gu}},
  \bibinfo {author} {\bibfnamefont {Z.-X.}\ \bibnamefont {Liu}}, \ and\
  \bibinfo {author} {\bibfnamefont {X.-G.}\ \bibnamefont {Wen}},\ }\href
  {\doibase 10.1103/physrevb.87.155114} {\bibfield  {journal} {\bibinfo
  {journal} {Phys. Rev. B}\ }\textbf {\bibinfo {volume} {87}},\ \bibinfo
  {pages} {155114} (\bibinfo {year} {2013})},\ \Eprint
  {http://arxiv.org/abs/arXiv:1106.4772} {arXiv:1106.4772} \BibitemShut
  {NoStop}%
\bibitem [{\citenamefont {Kapustin}\ and\ \citenamefont
  {Thorngren}(2014)}]{KT14030617}%
  \BibitemOpen
  \bibfield  {author} {\bibinfo {author} {\bibfnamefont {A.}~\bibnamefont
  {Kapustin}}\ and\ \bibinfo {author} {\bibfnamefont {R.}~\bibnamefont
  {Thorngren}},\ }\href {\doibase 10.1103/physrevlett.112.231602} {\bibfield
  {journal} {\bibinfo  {journal} {Phys. Rev. Lett.}\ }\textbf {\bibinfo
  {volume} {112}},\ \bibinfo {pages} {231602} (\bibinfo {year} {2014})},\
  \Eprint {http://arxiv.org/abs/arXiv:1403.0617} {arXiv:1403.0617} \BibitemShut
  {NoStop}%
\bibitem [{\citenamefont {Atiyah}(1988)}]{A8875}%
  \BibitemOpen
  \bibfield  {author} {\bibinfo {author} {\bibfnamefont {M.}~\bibnamefont
  {Atiyah}},\ }\href {\doibase 10.1007/bf02698547} {\bibfield  {journal}
  {\bibinfo  {journal} {Publications Math\'ematiques de l'Institut des Hautes
  Scientifiques}\ }\textbf {\bibinfo {volume} {68}},\ \bibinfo {pages} {175}
  (\bibinfo {year} {1988})}\BibitemShut {NoStop}%
\bibitem [{\citenamefont {{Freed}}(2014)}]{F1478}%
  \BibitemOpen
  \bibfield  {author} {\bibinfo {author} {\bibfnamefont {D.~S.}\ \bibnamefont
  {{Freed}}},\ }\href@noop {} {\  (\bibinfo {year} {2014})},\ \Eprint
  {http://arxiv.org/abs/arXiv:1406.7278} {arXiv:1406.7278} \BibitemShut
  {NoStop}%
\bibitem [{\citenamefont {Witten}(1989)}]{W8951}%
  \BibitemOpen
  \bibfield  {author} {\bibinfo {author} {\bibfnamefont {E.}~\bibnamefont
  {Witten}},\ }\href {\doibase 10.1007/bf01217730} {\bibfield  {journal}
  {\bibinfo  {journal} {Commun. Math. Phys.}\ }\textbf {\bibinfo {volume}
  {121}},\ \bibinfo {pages} {351} (\bibinfo {year} {1989})}\BibitemShut
  {NoStop}%
\bibitem [{\citenamefont {Fiorenza}\ and\ \citenamefont
  {Valentino}(2015)}]{FV14095723}%
  \BibitemOpen
  \bibfield  {author} {\bibinfo {author} {\bibfnamefont {D.}~\bibnamefont
  {Fiorenza}}\ and\ \bibinfo {author} {\bibfnamefont {A.}~\bibnamefont
  {Valentino}},\ }\href {\doibase 10.1007/s00220-015-2371-3} {\bibfield
  {journal} {\bibinfo  {journal} {Commun. Math. Phys.}\ }\textbf {\bibinfo
  {volume} {338}},\ \bibinfo {pages} {1043} (\bibinfo {year} {2015})},\ \Eprint
  {http://arxiv.org/abs/arXiv:1409.5723} {arXiv:1409.5723} \BibitemShut
  {NoStop}%
\bibitem [{\citenamefont {{Monnier}}(2015)}]{M14107442}%
  \BibitemOpen
  \bibfield  {author} {\bibinfo {author} {\bibfnamefont {S.}~\bibnamefont
  {{Monnier}}},\ }\href {\doibase 10.1007/s00220-015-2369-x} {\bibfield
  {journal} {\bibinfo  {journal} {Communications in Mathematical Physics}\
  }\textbf {\bibinfo {volume} {338}},\ \bibinfo {pages} {1327} (\bibinfo {year}
  {2015})},\ \Eprint {http://arxiv.org/abs/1410.7442} {arXiv:1410.7442}
  \BibitemShut {NoStop}%
\bibitem [{\citenamefont {{Wen}}\ and\ \citenamefont
  {{Wang}}(2018)}]{WW180109938}%
  \BibitemOpen
  \bibfield  {author} {\bibinfo {author} {\bibfnamefont {X.-G.}\ \bibnamefont
  {{Wen}}}\ and\ \bibinfo {author} {\bibfnamefont {Z.}~\bibnamefont {{Wang}}},\
  }\href@noop {} {\  (\bibinfo {year} {2018})},\ \Eprint
  {http://arxiv.org/abs/1801.09938} {arXiv:1801.09938} \BibitemShut {NoStop}%
\bibitem [{\citenamefont {Read}\ and\ \citenamefont {Sachdev}(1991)}]{RS9173}%
  \BibitemOpen
  \bibfield  {author} {\bibinfo {author} {\bibfnamefont {N.}~\bibnamefont
  {Read}}\ and\ \bibinfo {author} {\bibfnamefont {S.}~\bibnamefont {Sachdev}},\
  }\href@noop {} {\bibfield  {journal} {\bibinfo  {journal} {Phys. Rev. Lett.}\
  }\textbf {\bibinfo {volume} {66}},\ \bibinfo {pages} {1773} (\bibinfo {year}
  {1991})}\BibitemShut {NoStop}%
\bibitem [{\citenamefont {Wen}(1991)}]{W9164}%
  \BibitemOpen
  \bibfield  {author} {\bibinfo {author} {\bibfnamefont {X.-G.}\ \bibnamefont
  {Wen}},\ }\href {\doibase 10.1103/physrevb.44.2664} {\bibfield  {journal}
  {\bibinfo  {journal} {Phys. Rev. B}\ }\textbf {\bibinfo {volume} {44}},\
  \bibinfo {pages} {2664} (\bibinfo {year} {1991})}\BibitemShut {NoStop}%
\bibitem [{\citenamefont {Freedman}\ \emph {et~al.}(2004)\citenamefont
  {Freedman}, \citenamefont {Nayak}, \citenamefont {Shtengel}, \citenamefont
  {Walker},\ and\ \citenamefont {Wang}}]{FNS0428}%
  \BibitemOpen
  \bibfield  {author} {\bibinfo {author} {\bibfnamefont {M.}~\bibnamefont
  {Freedman}}, \bibinfo {author} {\bibfnamefont {C.}~\bibnamefont {Nayak}},
  \bibinfo {author} {\bibfnamefont {K.}~\bibnamefont {Shtengel}}, \bibinfo
  {author} {\bibfnamefont {K.}~\bibnamefont {Walker}}, \ and\ \bibinfo {author}
  {\bibfnamefont {Z.}~\bibnamefont {Wang}},\ }\href {\doibase
  10.1016/j.aop.2004.01.006} {\bibfield  {journal} {\bibinfo  {journal} {Ann.
  Phys.}\ }\textbf {\bibinfo {volume} {310}},\ \bibinfo {pages} {428} (\bibinfo
  {year} {2004})},\ \Eprint {http://arxiv.org/abs/cond-mat/0307511}
  {cond-mat/0307511} \BibitemShut {NoStop}%
\bibitem [{\citenamefont {Levin}\ and\ \citenamefont {Wen}(2005)}]{LW0510}%
  \BibitemOpen
  \bibfield  {author} {\bibinfo {author} {\bibfnamefont {M.~A.}\ \bibnamefont
  {Levin}}\ and\ \bibinfo {author} {\bibfnamefont {X.-G.}\ \bibnamefont
  {Wen}},\ }\href {\doibase 10.1103/physrevb.71.045110} {\bibfield  {journal}
  {\bibinfo  {journal} {Phys. Rev. B}\ }\textbf {\bibinfo {volume} {71}},\
  \bibinfo {pages} {045110} (\bibinfo {year} {2005})},\ \Eprint
  {http://arxiv.org/abs/cond-mat/0404617} {cond-mat/0404617} \BibitemShut
  {NoStop}%
\bibitem [{\citenamefont {Laughlin}(1983)}]{L8395}%
  \BibitemOpen
  \bibfield  {author} {\bibinfo {author} {\bibfnamefont {R.~B.}\ \bibnamefont
  {Laughlin}},\ }\href {\doibase 10.1103/physrevlett.50.1395} {\bibfield
  {journal} {\bibinfo  {journal} {Phys. Rev. Lett.}\ }\textbf {\bibinfo
  {volume} {50}},\ \bibinfo {pages} {1395} (\bibinfo {year}
  {1983})}\BibitemShut {NoStop}%
\bibitem [{Note1()}]{Note1}%
  \BibitemOpen
  \bibinfo {note} {The partition functions of a fermionic CFT on a $d+1$D
  lattice are MCG covariant in a special way, so that the sum of the partition
  functions are still invariant under the fermionic MCG. For example, in 1+1D,
  the fermionic MCG is generated by $T^2$ and $S$.}\BibitemShut {Stop}%
\bibitem [{\citenamefont {Keski-Vakkuri}\ and\ \citenamefont
  {Wen}(1993)}]{KW9327}%
  \BibitemOpen
  \bibfield  {author} {\bibinfo {author} {\bibfnamefont {E.}~\bibnamefont
  {Keski-Vakkuri}}\ and\ \bibinfo {author} {\bibfnamefont {X.-G.}\ \bibnamefont
  {Wen}},\ }\href {\doibase 10.1142/s0217979293003644} {\bibfield  {journal}
  {\bibinfo  {journal} {Int. J. Mod. Phys. B}\ }\textbf {\bibinfo {volume}
  {07}},\ \bibinfo {pages} {4227} (\bibinfo {year} {1993})}\BibitemShut
  {NoStop}%
\bibitem [{\citenamefont {Rowell}\ \emph {et~al.}(2009)\citenamefont {Rowell},
  \citenamefont {Stong},\ and\ \citenamefont {Wang}}]{RSW0777}%
  \BibitemOpen
  \bibfield  {author} {\bibinfo {author} {\bibfnamefont {E.}~\bibnamefont
  {Rowell}}, \bibinfo {author} {\bibfnamefont {R.}~\bibnamefont {Stong}}, \
  and\ \bibinfo {author} {\bibfnamefont {Z.}~\bibnamefont {Wang}},\ }\href
  {\doibase 10.1007/s00220-009-0908-z} {\bibfield  {journal} {\bibinfo
  {journal} {Commun. Math. Phys.}\ }\textbf {\bibinfo {volume} {292}},\
  \bibinfo {pages} {343} (\bibinfo {year} {2009})},\ \Eprint
  {http://arxiv.org/abs/arXiv:0712.1377} {arXiv:0712.1377} \BibitemShut
  {NoStop}%
\bibitem [{\citenamefont {Lan}\ \emph {et~al.}(2015)\citenamefont {Lan},
  \citenamefont {Wang},\ and\ \citenamefont {Wen}}]{LWW1414}%
  \BibitemOpen
  \bibfield  {author} {\bibinfo {author} {\bibfnamefont {T.}~\bibnamefont
  {Lan}}, \bibinfo {author} {\bibfnamefont {J.~C.}\ \bibnamefont {Wang}}, \
  and\ \bibinfo {author} {\bibfnamefont {X.-G.}\ \bibnamefont {Wen}},\ }\href
  {\doibase 10.1103/physrevlett.114.076402} {\bibfield  {journal} {\bibinfo
  {journal} {Phys. Rev. Lett.}\ }\textbf {\bibinfo {volume} {114}},\ \bibinfo
  {pages} {076402} (\bibinfo {year} {2015})},\ \Eprint
  {http://arxiv.org/abs/arXiv:1408.6514} {arXiv:1408.6514} \BibitemShut
  {NoStop}%
\bibitem [{\citenamefont {Lan}\ \emph {et~al.}()\citenamefont {Lan},
  \citenamefont {Wen}, \citenamefont {Kong},\ and\ \citenamefont
  {Wen}}]{dwall}%
  \BibitemOpen
  \bibfield  {author} {\bibinfo {author} {\bibfnamefont {T.}~\bibnamefont
  {Lan}}, \bibinfo {author} {\bibfnamefont {X.}~\bibnamefont {Wen}}, \bibinfo
  {author} {\bibfnamefont {L.}~\bibnamefont {Kong}}, \ and\ \bibinfo {author}
  {\bibfnamefont {X.-G.}\ \bibnamefont {Wen}},\ }\href@noop {} {\bibinfo
  {journal} {To appear}\ }\BibitemShut {NoStop}%
\bibitem [{\citenamefont {{Kong}}\ and\ \citenamefont
  {{Zheng}}(2018)}]{KZ170501087}%
  \BibitemOpen
\bibfield  {journal} {  }\bibfield  {author} {\bibinfo {author} {\bibfnamefont
  {L.}~\bibnamefont {{Kong}}}\ and\ \bibinfo {author} {\bibfnamefont
  {H.}~\bibnamefont {{Zheng}}},\ }\href {\doibase
  10.1016/j.nuclphysb.2017.12.007} {\bibfield  {journal} {\bibinfo  {journal}
  {Nuclear Physics B}\ }\textbf {\bibinfo {volume} {927}},\ \bibinfo {pages}
  {140} (\bibinfo {year} {2018})},\ \Eprint {http://arxiv.org/abs/1705.01087}
  {arXiv:1705.01087} \BibitemShut {NoStop}%
\bibitem [{\citenamefont {{Chen}}\ \emph {et~al.}(2019)\citenamefont {{Chen}},
  \citenamefont {{Jian}}, \citenamefont {{Kong}}, \citenamefont {{You}},\ and\
  \citenamefont {{Zheng}}}]{CZ190312334}%
  \BibitemOpen
  \bibfield  {author} {\bibinfo {author} {\bibfnamefont {W.-Q.}\ \bibnamefont
  {{Chen}}}, \bibinfo {author} {\bibfnamefont {C.-M.}\ \bibnamefont {{Jian}}},
  \bibinfo {author} {\bibfnamefont {L.}~\bibnamefont {{Kong}}}, \bibinfo
  {author} {\bibfnamefont {Y.-Z.}\ \bibnamefont {{You}}}, \ and\ \bibinfo
  {author} {\bibfnamefont {H.}~\bibnamefont {{Zheng}}},\ }\href@noop {} {\ ,\
  \bibinfo {pages} {arXiv:1903.12334} (\bibinfo {year} {2019})},\ \Eprint
  {http://arxiv.org/abs/1903.12334} {arXiv:1903.12334} \BibitemShut {NoStop}%
\bibitem [{\citenamefont {{Lin}}\ and\ \citenamefont
  {{Shao}}(2019)}]{LS190404833}%
  \BibitemOpen
  \bibfield  {author} {\bibinfo {author} {\bibfnamefont {Y.-H.}\ \bibnamefont
  {{Lin}}}\ and\ \bibinfo {author} {\bibfnamefont {S.-H.}\ \bibnamefont
  {{Shao}}},\ }\href@noop {} {\  (\bibinfo {year} {2019})},\ \Eprint
  {http://arxiv.org/abs/1904.04833} {arXiv:1904.04833} \BibitemShut {NoStop}%
\bibitem [{\citenamefont {Ryu}\ and\ \citenamefont {Zhang}(2012)}]{RZ1232}%
  \BibitemOpen
  \bibfield  {author} {\bibinfo {author} {\bibfnamefont {S.}~\bibnamefont
  {Ryu}}\ and\ \bibinfo {author} {\bibfnamefont {S.-C.}\ \bibnamefont
  {Zhang}},\ }\href {\doibase 10.1103/physrevb.85.245132} {\bibfield  {journal}
  {\bibinfo  {journal} {Phys. Rev. B}\ }\textbf {\bibinfo {volume} {85}},\
  \bibinfo {pages} {245132} (\bibinfo {year} {2012})}\BibitemShut {NoStop}%
\bibitem [{\citenamefont {{Tiwari}}\ \emph {et~al.}(2018)\citenamefont
  {{Tiwari}}, \citenamefont {{Chen}}, \citenamefont {{Shiozaki}},\ and\
  \citenamefont {{Ryu}}}]{TR171004730}%
  \BibitemOpen
  \bibfield  {author} {\bibinfo {author} {\bibfnamefont {A.}~\bibnamefont
  {{Tiwari}}}, \bibinfo {author} {\bibfnamefont {X.}~\bibnamefont {{Chen}}},
  \bibinfo {author} {\bibfnamefont {K.}~\bibnamefont {{Shiozaki}}}, \ and\
  \bibinfo {author} {\bibfnamefont {S.}~\bibnamefont {{Ryu}}},\ }\href
  {\doibase 10.1103/PhysRevB.97.245133} {\bibfield  {journal} {\bibinfo
  {journal} {\prb}\ }\textbf {\bibinfo {volume} {97}},\ \bibinfo {pages}
  {245133} (\bibinfo {year} {2018})},\ \Eprint
  {http://arxiv.org/abs/1710.04730} {arXiv:1710.04730} \BibitemShut {NoStop}%
\bibitem [{\citenamefont {Ji}\ and\ \citenamefont {Wen}()}]{STanomaly}%
  \BibitemOpen
  \bibfield  {author} {\bibinfo {author} {\bibfnamefont {W.}~\bibnamefont
  {Ji}}\ and\ \bibinfo {author} {\bibfnamefont {X.-G.}\ \bibnamefont {Wen}},\
  }\href@noop {} {\bibinfo  {journal} {To appear}\ }\BibitemShut {NoStop}%
\bibitem [{\citenamefont {Chamon}(2005)}]{C0502}%
  \BibitemOpen
\bibfield  {journal} {  }\bibfield  {author} {\bibinfo {author} {\bibfnamefont
  {C.}~\bibnamefont {Chamon}},\ }\href {\doibase 10.1103/physrevlett.94.040402}
  {\bibfield  {journal} {\bibinfo  {journal} {Phys. Rev. Lett.}\ }\textbf
  {\bibinfo {volume} {94}},\ \bibinfo {pages} {040402} (\bibinfo {year}
  {2005})}\BibitemShut {NoStop}%
\bibitem [{\citenamefont {Haah}(2011)}]{H11011962}%
  \BibitemOpen
  \bibfield  {author} {\bibinfo {author} {\bibfnamefont {J.}~\bibnamefont
  {Haah}},\ }\href {\doibase 10.1103/physreva.83.042330} {\bibfield  {journal}
  {\bibinfo  {journal} {Phys. Rev. A}\ }\textbf {\bibinfo {volume} {83}},\
  \bibinfo {pages} {042330} (\bibinfo {year} {2011})},\ \Eprint
  {http://arxiv.org/abs/arXiv:1101.1962} {arXiv:1101.1962} \BibitemShut
  {NoStop}%
\bibitem [{\citenamefont {Zeng}\ and\ \citenamefont {Wen}(2015)}]{ZW1490}%
  \BibitemOpen
  \bibfield  {author} {\bibinfo {author} {\bibfnamefont {B.}~\bibnamefont
  {Zeng}}\ and\ \bibinfo {author} {\bibfnamefont {X.-G.}\ \bibnamefont {Wen}},\
  }\href {\doibase 10.1103/physrevb.91.125121} {\bibfield  {journal} {\bibinfo
  {journal} {Phys. Rev. B}\ }\textbf {\bibinfo {volume} {91}},\ \bibinfo
  {pages} {125121} (\bibinfo {year} {2015})},\ \Eprint
  {http://arxiv.org/abs/arXiv:1406.5090} {arXiv:1406.5090} \BibitemShut
  {NoStop}%
\bibitem [{\citenamefont {{Shirley}}\ \emph
  {et~al.}(2018{\natexlab{a}})\citenamefont {{Shirley}}, \citenamefont
  {{Slagle}}, \citenamefont {{Wang}},\ and\ \citenamefont
  {{Chen}}}]{SC171205892}%
  \BibitemOpen
  \bibfield  {author} {\bibinfo {author} {\bibfnamefont {W.}~\bibnamefont
  {{Shirley}}}, \bibinfo {author} {\bibfnamefont {K.}~\bibnamefont {{Slagle}}},
  \bibinfo {author} {\bibfnamefont {Z.}~\bibnamefont {{Wang}}}, \ and\ \bibinfo
  {author} {\bibfnamefont {X.}~\bibnamefont {{Chen}}},\ }\href {\doibase
  10.1103/PhysRevX.8.031051} {\bibfield  {journal} {\bibinfo  {journal}
  {Physical Review X}\ }\textbf {\bibinfo {volume} {8}},\ \bibinfo {pages}
  {031051} (\bibinfo {year} {2018}{\natexlab{a}})},\ \Eprint
  {http://arxiv.org/abs/1712.05892} {arXiv:1712.05892} \BibitemShut {NoStop}%
\bibitem [{\citenamefont {{Shirley}}\ \emph
  {et~al.}(2018{\natexlab{b}})\citenamefont {{Shirley}}, \citenamefont
  {{Slagle}},\ and\ \citenamefont {{Chen}}}]{SC180310426}%
  \BibitemOpen
  \bibfield  {author} {\bibinfo {author} {\bibfnamefont {W.}~\bibnamefont
  {{Shirley}}}, \bibinfo {author} {\bibfnamefont {K.}~\bibnamefont {{Slagle}}},
  \ and\ \bibinfo {author} {\bibfnamefont {X.}~\bibnamefont {{Chen}}},\
  }\href@noop {} {\ ,\ \bibinfo {pages} {arXiv:1803.10426} (\bibinfo {year}
  {2018}{\natexlab{b}})},\ \Eprint {http://arxiv.org/abs/1803.10426}
  {arXiv:1803.10426} \BibitemShut {NoStop}%
\bibitem [{\citenamefont {Wen}(2014)}]{W1447}%
  \BibitemOpen
  \bibfield  {author} {\bibinfo {author} {\bibfnamefont {X.-G.}\ \bibnamefont
  {Wen}},\ }\href {\doibase 10.1103/physrevb.89.035147} {\bibfield  {journal}
  {\bibinfo  {journal} {Phys. Rev. B}\ }\textbf {\bibinfo {volume} {89}},\
  \bibinfo {pages} {035147} (\bibinfo {year} {2014})},\ \Eprint
  {http://arxiv.org/abs/arXiv:1301.7675} {arXiv:1301.7675} \BibitemShut
  {NoStop}%
\bibitem [{\citenamefont {Hung}\ and\ \citenamefont {Wen}(2014)}]{HW1339}%
  \BibitemOpen
  \bibfield  {author} {\bibinfo {author} {\bibfnamefont {L.-Y.}\ \bibnamefont
  {Hung}}\ and\ \bibinfo {author} {\bibfnamefont {X.-G.}\ \bibnamefont {Wen}},\
  }\href {\doibase 10.1103/physrevb.89.075121} {\bibfield  {journal} {\bibinfo
  {journal} {Phys. Rev. B}\ }\textbf {\bibinfo {volume} {89}},\ \bibinfo
  {pages} {075121} (\bibinfo {year} {2014})},\ \Eprint
  {http://arxiv.org/abs/arXiv:1311.5539} {arXiv:1311.5539} \BibitemShut
  {NoStop}%
\bibitem [{\citenamefont {Kapustin}(2014)}]{K1459}%
  \BibitemOpen
  \bibfield  {author} {\bibinfo {author} {\bibfnamefont {A.}~\bibnamefont
  {Kapustin}},\ }\href@noop {} {\  (\bibinfo {year} {2014})},\ \Eprint
  {http://arxiv.org/abs/arXiv:1404.6659} {arXiv:1404.6659} \BibitemShut
  {NoStop}%
\bibitem [{\citenamefont {Wen}\ and\ \citenamefont {Zee}(1992)}]{WZ9290}%
  \BibitemOpen
  \bibfield  {author} {\bibinfo {author} {\bibfnamefont {X.-G.}\ \bibnamefont
  {Wen}}\ and\ \bibinfo {author} {\bibfnamefont {A.}~\bibnamefont {Zee}},\
  }\href {\doibase 10.1103/physrevb.46.2290} {\bibfield  {journal} {\bibinfo
  {journal} {Phys. Rev. B}\ }\textbf {\bibinfo {volume} {46}},\ \bibinfo
  {pages} {2290} (\bibinfo {year} {1992})}\BibitemShut {NoStop}%
\bibitem [{\citenamefont {Wen}(1995)}]{W9505}%
  \BibitemOpen
  \bibfield  {author} {\bibinfo {author} {\bibfnamefont {X.-G.}\ \bibnamefont
  {Wen}},\ }\href {\doibase 10.1080/00018739500101566} {\bibfield  {journal}
  {\bibinfo  {journal} {Adv. Phys.}\ }\textbf {\bibinfo {volume} {44}},\
  \bibinfo {pages} {405} (\bibinfo {year} {1995})},\ \Eprint
  {http://arxiv.org/abs/cond-mat/9506066} {cond-mat/9506066} \BibitemShut
  {NoStop}%
\bibitem [{\citenamefont {{Yang}}\ \emph {et~al.}(2014)\citenamefont {{Yang}},
  \citenamefont {{Lehman}}, \citenamefont {{Poilblanc}}, \citenamefont {{Van
  Acoleyen}}, \citenamefont {{Verstraete}}, \citenamefont {{Cirac}},\ and\
  \citenamefont {{Schuch}}}]{YS13094596}%
  \BibitemOpen
  \bibfield  {author} {\bibinfo {author} {\bibfnamefont {S.}~\bibnamefont
  {{Yang}}}, \bibinfo {author} {\bibfnamefont {L.}~\bibnamefont {{Lehman}}},
  \bibinfo {author} {\bibfnamefont {D.}~\bibnamefont {{Poilblanc}}}, \bibinfo
  {author} {\bibfnamefont {K.}~\bibnamefont {{Van Acoleyen}}}, \bibinfo
  {author} {\bibfnamefont {F.}~\bibnamefont {{Verstraete}}}, \bibinfo {author}
  {\bibfnamefont {J.~I.}\ \bibnamefont {{Cirac}}}, \ and\ \bibinfo {author}
  {\bibfnamefont {N.}~\bibnamefont {{Schuch}}},\ }\href {\doibase
  10.1103/PhysRevLett.112.036402} {\bibfield  {journal} {\bibinfo  {journal}
  {\prl}\ }\textbf {\bibinfo {volume} {112}},\ \bibinfo {pages} {036402}
  (\bibinfo {year} {2014})},\ \Eprint {http://arxiv.org/abs/1309.4596}
  {arXiv:1309.4596} \BibitemShut {NoStop}%
\bibitem [{\citenamefont {Chen}\ and\ \citenamefont {Wen}(2012)}]{CW1235}%
  \BibitemOpen
  \bibfield  {author} {\bibinfo {author} {\bibfnamefont {X.}~\bibnamefont
  {Chen}}\ and\ \bibinfo {author} {\bibfnamefont {X.-G.}\ \bibnamefont {Wen}},\
  }\href@noop {} {\bibfield  {journal} {\bibinfo  {journal} {Phys. Rev. B}\
  }\textbf {\bibinfo {volume} {86}},\ \bibinfo {pages} {235135} (\bibinfo
  {year} {2012})},\ \Eprint {http://arxiv.org/abs/arXiv:1206.3117}
  {arXiv:1206.3117} \BibitemShut {NoStop}%
\bibitem [{\citenamefont {Francesco}\ \emph {et~al.}(2012)\citenamefont
  {Francesco}, \citenamefont {Mathieu},\ and\ \citenamefont
  {S{\'e}n{\'e}chal}}]{francesco2012conformal}%
  \BibitemOpen
  \bibfield  {author} {\bibinfo {author} {\bibfnamefont {P.}~\bibnamefont
  {Francesco}}, \bibinfo {author} {\bibfnamefont {P.}~\bibnamefont {Mathieu}},
  \ and\ \bibinfo {author} {\bibfnamefont {D.}~\bibnamefont
  {S{\'e}n{\'e}chal}},\ }\href@noop {} {\emph {\bibinfo {title} {Conformal
  field theory}}}\ (\bibinfo  {publisher} {Springer Science \& Business
  Media},\ \bibinfo {year} {2012})\BibitemShut {NoStop}%
\bibitem [{\citenamefont {Cardy}(1996)}]{cardy1996scaling}%
  \BibitemOpen
  \bibfield  {author} {\bibinfo {author} {\bibfnamefont {J.}~\bibnamefont
  {Cardy}},\ }\href@noop {} {\emph {\bibinfo {title} {Scaling and
  renormalization in statistical physics}}},\ Vol.~\bibinfo {volume} {5}\
  (\bibinfo  {publisher} {Cambridge university press},\ \bibinfo {year}
  {1996})\BibitemShut {NoStop}%
\bibitem [{\citenamefont {Levin}\ and\ \citenamefont {Gu}(2012)}]{LG1209}%
  \BibitemOpen
  \bibfield  {author} {\bibinfo {author} {\bibfnamefont {M.}~\bibnamefont
  {Levin}}\ and\ \bibinfo {author} {\bibfnamefont {Z.-C.}\ \bibnamefont {Gu}},\
  }\href {\doibase 10.1103/physrevb.86.115109} {\bibfield  {journal} {\bibinfo
  {journal} {Phys. Rev. B}\ }\textbf {\bibinfo {volume} {86}},\ \bibinfo
  {pages} {115109} (\bibinfo {year} {2012})},\ \Eprint
  {http://arxiv.org/abs/arXiv:1202.3120} {arXiv:1202.3120} \BibitemShut
  {NoStop}%
\bibitem [{\citenamefont {Mathur}\ \emph {et~al.}(1989)\citenamefont {Mathur},
  \citenamefont {Mukhi},\ and\ \citenamefont {Sen}}]{mathur1989reconstruction}%
  \BibitemOpen
  \bibfield  {author} {\bibinfo {author} {\bibfnamefont {S.~D.}\ \bibnamefont
  {Mathur}}, \bibinfo {author} {\bibfnamefont {S.}~\bibnamefont {Mukhi}}, \
  and\ \bibinfo {author} {\bibfnamefont {A.}~\bibnamefont {Sen}},\ }\href@noop
  {} {\bibfield  {journal} {\bibinfo  {journal} {Nuclear Physics B}\ }\textbf
  {\bibinfo {volume} {318}},\ \bibinfo {pages} {483} (\bibinfo {year}
  {1989})}\BibitemShut {NoStop}%
\bibitem [{\citenamefont {Costantino}(2005)}]{C0527}%
  \BibitemOpen
  \bibfield  {author} {\bibinfo {author} {\bibfnamefont {F.}~\bibnamefont
  {Costantino}},\ }\href {\doibase 10.1007/s00209-005-0810-0} {\bibfield
  {journal} {\bibinfo  {journal} {Math. Z.}\ }\textbf {\bibinfo {volume}
  {251}},\ \bibinfo {pages} {427} (\bibinfo {year} {2005})},\ \Eprint
  {http://arxiv.org/abs/math/0403014} {math/0403014} \BibitemShut {NoStop}%
\bibitem [{\citenamefont {Levin}\ and\ \citenamefont {Nave}(2007)}]{LN0701}%
  \BibitemOpen
  \bibfield  {author} {\bibinfo {author} {\bibfnamefont {M.}~\bibnamefont
  {Levin}}\ and\ \bibinfo {author} {\bibfnamefont {C.~P.}\ \bibnamefont
  {Nave}},\ }\href {\doibase 10.1103/physrevlett.99.120601} {\bibfield
  {journal} {\bibinfo  {journal} {Phys. Rev. Lett.}\ }\textbf {\bibinfo
  {volume} {99}},\ \bibinfo {pages} {120601} (\bibinfo {year}
  {2007})}\BibitemShut {NoStop}%
\bibitem [{\citenamefont {{Zhu}}\ \emph {et~al.}(2018)\citenamefont {{Zhu}},
  \citenamefont {{Lan}},\ and\ \citenamefont {{Wen}}}]{ZW180809394}%
  \BibitemOpen
  \bibfield  {author} {\bibinfo {author} {\bibfnamefont {C.}~\bibnamefont
  {{Zhu}}}, \bibinfo {author} {\bibfnamefont {T.}~\bibnamefont {{Lan}}}, \ and\
  \bibinfo {author} {\bibfnamefont {X.-G.}\ \bibnamefont {{Wen}}},\ }\href@noop
  {} {\  (\bibinfo {year} {2018})},\ \Eprint {http://arxiv.org/abs/1808.09394}
  {arXiv:1808.09394} \BibitemShut {NoStop}%
\end{thebibliography}%

\end{document}